\documentclass[journal]{new-aiaa}
\usepackage{new-mysymb}
\usepackage{longtable}
\usepackage{multirow}
\usepackage{pbox}
\usepackage[hyphens]{url}
\usepackage{xcolor}
\usepackage{soul}
\usepackage{nomencl}
\usepackage{etoolbox}
\usepackage[ruled,vlined,linesnumbered]{algorithm2e}

\renewcommand\nomgroup[1]{%
   \item[\bfseries
   \ifstrequal{#1}{A}{Subscripts / Superscripts}]}
   
\makenomenclature
\printnomenclature

\newcommand{\bz}{\mathbf{z}}
\newcommand{\blambda}{\boldsymbol{\lambda}}

\newcommand{\bTheta}{\boldsymbol{\Theta}}

\newcommand{\bbeta}{\boldsymbol{\beta}}

\newcommand\cD{\mathcal{D}}

\newcommand{\barvy}{\bar{\mathbf{y}}}
\newcommand{\dotbarvy}{\dot{\bar{\mathbf{y}}}}

\soulregister\cite7
\soulregister\ref7
\newcommand\CORRECTa[1]{{#1}}

\title{Physics-Infused Reduced Order Modeling of Aerothermal Loads for Hypersonic Aerothermoelastic Analysis}

\author{Carlos Vargas Venegas\footnote{Graduate Student Research Assistant, Student Member AIAA}}
\author{Daning Huang\footnote{Assistant Professor, Member AIAA}}
\affil{Department of Aerospace Engineering, The Pennsylvania State University, University Park, PA, 16802}

\begin{document}
\maketitle

\begin{abstract}
This paper presents a novel physics-infused reduced-order modeling (PIROM) methodology for efficient and accurate modeling of non-linear dynamical systems. The PIROM consists of a \CORRECTa{physics}-based analytical component that represents the known physical processes, and a data-driven dynamical component that represents the unknown physical processes. The PIROM is applied to the aerothermal load modeling for hypersonic aerothermoelastic (ATE) analysis and is found to accelerate the ATE simulations by two-three orders of magnitude while maintaining an accuracy comparable to high-fidelity solutions based on computational fluid dynamics (CFD). Moreover, the PIROM-based solver is benchmarked against the conventional POD-kriging surrogate model, and is found to significantly outperform the \CORRECTa{accuracy, generalizability and sampling efficiency} of the latter in a wide range of operating conditions and in the presence of complex structural boundary conditions. Finally, the PIROM-based ATE solver is demonstrated by a parametric study on the effects of boundary conditions and rib-supports on the ATE response of a compliant and heat-conducting panel structure. The results not only reveal the dramatic snap-through behavior with respect to spring constraints of boundary conditions, but also demonstrates the potential of PIROM to facilitate the rapid and accurate design and optimization of multi-disciplinary systems such as hypersonic structures.
\end{abstract}

\begin{center}
    \bfseries List of Symbols
\end{center}

\begin{longtable}{p{3.5cm}p{13cm}}
    $\mathbf{A}$  & Data-driven system matrix \\
    $C_{f}$       & Skin-friction coefficient \\
    $\mathbf{c}$  & Auxiliary variables \\
    $\cD$         & Training dataset \\
    $\mathbf{d}$  & Kriging input vector \\
    $\mathbf{f}$  & Model system dynamics \\
    $\mathbf{g}$  & Auxiliary variable equations \\
    $H$           & Boundary layer shape factor \\
    $h$           & Panel thickness \\
    $\mathbf{h}$  & Output equations \\
    $\mathcal{J}$ & Cost function \\
    $k_p$         & Boundary layer wall and edge pressure ratio\\
    $L$           & Geometrical length \\
    $M$           & Mach number \\
    $\mathbf{M}$  & Model system matrix \\
    $N_s$         & Number of high-fidelity solutions \\
    $P$           & Pressure \\
    $Pr$          & Prandtl number \\
    $\dot{q}$     & Heat flux \\
    $r_f$         & Recovery factor \\
    $Re$          & Reynold's number \\
    $St$          & Stanton number \\
    $T$           & Temperature \\
    $\vu$         & Input variables \\
    $x,z$         & Coordinate axis \\
    $\mathbf{z}$  & Output variables \\
    $\vtb$        & Augmentation variables \\
    $\vtG$        & Data-driven dynamics\\
    $\delta^*$    & Boundary layer displacement thickness \\
    $\gamma$      & Specific heat ratio \\
    $\vtQ$        & Learnable parameters for data-driven models \\
    $\vtl$        & Adjoint variable \\
    $\mu$         & Dynamic viscosity \\
    $\rho$        & Density \\
    $\vtt$        & System parameters \\
    $\Box_e$      & Quantities at the edge of the boundary layer \\
    $\Box_{GP}$   & Quantities computed using a GP model\\
    $\Box_{HF}$   & Quantities related to CFD-based solutions \\
    $\Box_{spl}$  & Spline-interpolated quantities \\
    \CORRECTa{$\Box_{r}$}    & \CORRECTa{Quantities evaluated at reference conditions} \\
    $\Box_w$      & Quantities at the panel surface \\
    $\Box_\infty$ & Quantities related to far freestream \\
\addtocounter{table}{-1}
\end{longtable}

\section{Introduction} \label{sec:introduction}

During atmospheric flight, hypersonic vehicles create strong vehicle-atmosphere interactions that induce complex and highly-energetic flow physics. The flow physics exhibits rich and complex characteristics such as thin shock layers, entropy layers, viscous interactions, and low-density high-temperature flows, which are not present in subsonic or low supersonic conditions \cite{Huang2019b}. The interactions between the flow and the vehicle result in highly nonlinear coupling between aerodynamics, structures, and thermal responses, producing a fluid-thermal-structural interaction (FTSI) problem denoted by the term hypersonic \textit{aerothermoelasticity} (ATE). The prediction of the ATE response of hypersonic structures is critical for proper vehicle design, and requires extensive experimentation and computational analysis \cite{Crowell2010,Crowell2011}. Due to the impracticality of ground-based wind tunnel facilities \cite{Bertin2006}, most of the analysis and design of hypersonic structures has remained in the computational domain \cite{Tzong2010,McNamara2010,McNamara2011}.

Reliable computational modeling of hypersonic ATE requires high-fidelity structural, thermal, and aerodynamic solvers, as well as robust coupling of these solvers in a computational framework. With the rapidly growing computational power, the availability of high-fidelity data and mathematical modeling tools have enabled unprecedented computational capabilities in the multi-disciplinary solvers \cite{Tzong2010,McNamara2011,Quiroz2012,Bowcutt2018}. Particularly, numerical approaches such as computational fluid dynamics (CFD) provided unprecedented fidelity in the prediction of the hypersonic flow, and enabled the identification of flow phenomena such as dissociation, chemically reacting flows, and viscous interactions. Despite the advances in modeling and computational power discussed above, hypersonic ATE simulations of ever-high fidelity are still computationally intractable at the time-scale of a complete flight trajectory.

To overcome the computational shortcomings of CFD-based hypersonic ATE analysis and design, reduced-order models (ROMs) have been introduced into hypersonic ATE computational frameworks as substitutes to the CFD-based solver. A commonly used approach, is the aerodynamic surrogate based on proper-orthogonal decomposition (POD) \cite{Berkooz1993} and kriging \cite{Rasmussen2006} (POD-kriging), and has demonstrated reasonably accurate and efficient aerothermal load modeling for hypersonic ATE analysis \cite{Crowell2011,Crowell2012,Crowell2014,Huang2019b}. However, this surrogate model suffers from two fundamental limitations: generalization and characterization \cite{Huang2019,Huang2020b}. On the generalization, the POD-kriging surrogate cannot be generalized for configurations, flow conditions, and thermoelastic responses, i.e. the structural deformation and temperature distribution, that are different from those used in the training data samples. On the characterization, due to the curse of dimensionality, i.e. the required number of training samples grows rapidly with respect to the number of inputs, it soon becomes impractical to generate aerothermal surrogates for complex hypersonic vehicle configurations. 

Several efforts have been conducted to alleviate the generalization and characterization limitations of the POD-kriging surrogate \cite{Crowell2011,Crowell2014,Rokita2018,Sadagopan2020,Sadagopan2021}. In Refs. \cite{Rokita2018,Sadagopan2021}, multi-fidelity kriging models termed POD-coKriging and $\text{M}^2$GPR, respectively, were developed to predict aerothermal loads over deforming structures subjected to hypersonic flow. The principal idea is to train a kriging model using a large number of low-fidelity samples, and construct an additional difference kriging model that exploits the correlation of the low-fidelity model and high-fidelity outputs. The multi-fidelity kriging models have demonstrated improved accuracy for undersampled datasets and extended abilities to extrapolation relative to the conventional POD-kriging surrogate. However, these approaches may be prone to inaccuracies when applied to high-dimensional and highly nonlinear systems due to model simplifications such as diagonal scale factor assumptions in the POD-coKriging formulation \cite{Rokita2018}, and uncorrelation assumptions for the difference and low-fidelity kriging models in the $\text{M}^2$GPR formulation \cite{Sadagopan2021}.

Other studies \cite{Crowell2011,Crowell2014,Huang2019b} have focused on developing correction techniques for the aerothermal load predictions in hypersonic ATE analysis. These approaches alleviate the generalization issues by introducing analytical or semi-empirical formulations that compensate for varying temperature and structural distributions, varying flight conditions, and geometric scales \cite{Huang2019b,Sadagopan2020}. However, the correction techniques usually work in the vicinity of the reference solutions and geometrically similar configurations, and are not sufficiently general for aerothermal load prediction for \textit{arbitrary} configurations and thermoelastic responses.

In addition to the modeling approaches discussed above, there have been efforts for aerothermal load prediction using \CORRECTa{first-order} physics-based analytical models. One classical example is the turbulent viscous-inviscid interaction (TVI) model that was originally developed in Refs. \cite{Stollery1969,Stollery1973}.  The TVI model has mostly been used for steady aerodynamic load predictions of deformed structures subjected to high-speed flow \cite{Brouwer2016,Brouwer2020}. In Ref. \cite{Brouwer2016}, the TVI model is formulated using semi-empirical expressions for the boundary layer shape factor and skin-friction coefficient derived from CFD-based solutions, and is coupled to inviscid pressure models such as piston theory \cite{Meijer2016,Meijer2018}. Pressure load predictions showed considerable improvements in the average error relative to localized piston theory and free vibration mode methods. In Ref. \cite{Brouwer2020}, the TVI model is used to identify a set of parameters that impact the effects of general static deformations on aerodynamic load predictions. Then, a pointwise data-driven model is trained from CFD-based simulations over prescribed deformations using the TVI-identified input space to predict the pressure distributions over an arbitrarily deformed orthogrid panel structure. Pressure loads correlated well with steady CFD-based predictions, but suffered from errors at the trailing edge of the panel due to required integrations over the spatial domain in the identified input space. The TVI model has also found successful applications in aerothermal load modeling for arbitrarily deformed two-dimensional skin panel structures \cite{Venegas2021,Venegas2022} when coupled to the Eckert's reference enthalpy method \cite{Eckert1956,Eckert1960} for heat flux predictions \cite{Crowell2011,McNamara2010,Anderson2006}. In general, despite the efficiency and generalizability of TVI, the model underperforms quantitatively due to simplifications and assumptions in its formulation, hence prohibiting its practical use to high-fidelity analysis and design of hypersonic structures.

A promising alternative methodology for the aerothermal load modeling is the data-driven model calibration method known as field inversion and machine learning (FIML) \cite{Duraisamy2013,Duraisamy2016,Holland2019,Holland2019a,Holland2019b}. The objective of FIML is to correct the inaccuracies of a low-order analytical model using high-fidelity data. The FIML performs an intrusive modification of the low-order model equations by introducing an unknown \textit{augmentation} function that intends to compensate for the missing physics in the low-order model. Then, field inversion and machine learning stages are conducted to extract, learn, and represent the augmentation field along the computational domain using a data-driven model. The FIML approach has demonstrated excellent model improvement capabilities for fluid problems involving complex geometries \cite{Holland2019a,Holland2019b,Duraisamy2016}. In Refs. \cite{Holland2019a,Holland2019b}, the FIML methodology was found to provide the necessary corrections to the turbulence production term in the Spalart-Allmaras turbulence model for flow past airfoils at high angles of attack. This was achieved by utilizing a neural network (NN) to learn the augmentation fields extracted during the field inversion stage, and then using the trained NN to generate predictions under arbitrary system configurations to improve the turbulence production term. In Ref. \cite{Duraisamy2016}, a Bayesian FIML formulation identified the unknown augmentation field given only a handful of data points along the computational domain for turbulence modeling applications. The augmentation field was used in the low-fidelity transport equations, and enabled high correlation of the posterior model predictions with high-fidelity data for problems involving channel flows, shock-boundary layer interactions, and flow with curvature and separation \cite{Duraisamy2016}. In general, the FIML methodology is effective in extracting the spatio-temporal augmentation fields in a computational domain, and using these to correct the low-order models.

In this work, the \textit{physics-infused reduced order modeling} (PIROM) methodology is presented based on preliminary studies \cite{Venegas2021,Venegas2022}, and is adopted for the creation of a robust, accurate, and efficient ROMs with applications to aerothermal load modeling in hypersonic ATE applications. Following the principles of the FIML approach, the PIROM methodology explicitly couples a known low-fidelity physics-based differential-algebraic model, with an \textit{a priori} unknown data-driven model. However, the PIROM is developed as a superset of the FIML formulation; it extends the algebraic data-driven augmentation to a differential form that is potentially more suitable for a dynamical system and offers superior expressibility for the functional form of the augmentation terms.

The PIROM-based aerothermodynamic solver is for the first time incorporated into the well-verified eXtended HYPersonic ATE (HYPATE-X) \cite{Huang2017a,Huang2019b} computational framework, illustrated in Fig. \ref{figs/sec1/hypate-code-structure.PNG}, and is used to conduct high-fidelity hypersonic ATE analysis of a two-dimensional compliant and heat-conducting structure subjected to hypersonic flow under various structural boundary condition configurations. Conventional hypersonic ATE analysis has been widely performed for two-dimensional simply-supported structures undergoing cylindrical bending \cite{McNamara2010,Culler2010,Brouwer2018,Huang2019}, while others studies have considered built-up configurations such as clamped panel structures with stringer supports along the transverse and chordwise directions \cite{Quiroz2012,Brouwer2020}. The present study extends the hypersonic ATE analysis by subjecting the aerothermal models to panel structures under clamped, spring, and rib-supported boundary conditions; these configurations effectively render the structures as different geometrical configurations for which conventional ROMs, e.g., the POD-kriging surrogate, cannot be easily applied.

\insertfig{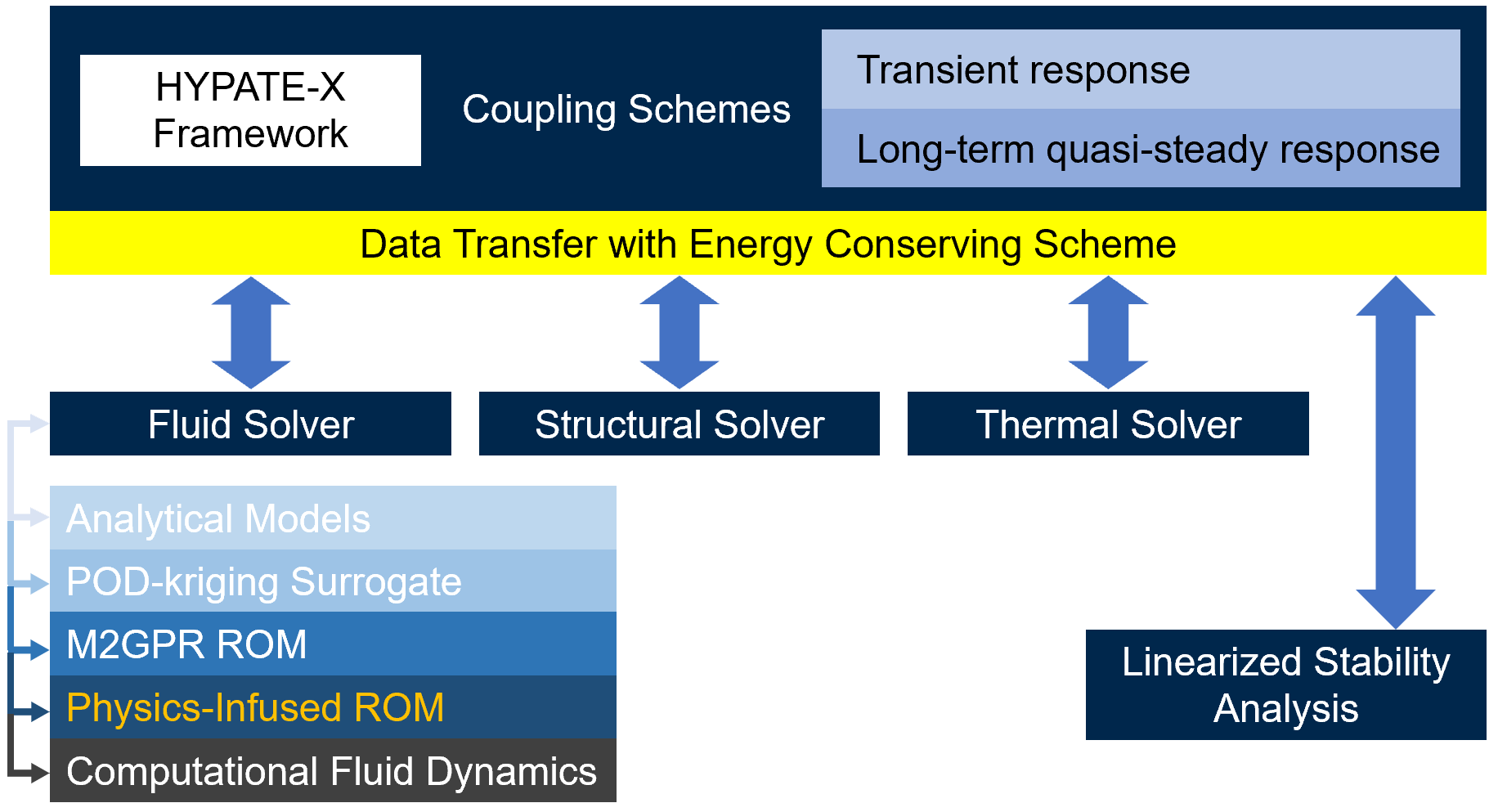}{0.50}{HYPATE-X framework with the new PIROM module in yellow.}

In sum, the objectives of this study are to,
\begin{compactenum}
    \item Present and apply the PIROM methodology to construct an aerothermal model for hypersonic ATE analysis.
    \item Benchmark the PIROM-based aerothermal model against the conventional \CORRECTa{method, and highlight the advantages of PIROM, especially the generalizability and sampling efficiency.}
    \item Demonstrate the efficiency, accuracy, and generalizability of PIROM in coupled ATE analysis.
    \item Examine the effects of structural boundary conditions on the transient ATE response of a typical high-speed panel structure.
\end{compactenum}


\section{Physics-Infused Reduced-Order Modeling for Aerothermal Loads}\label{sec:pirom}
This section presents the PIROM formulation in the context of aerothermal load prediction. The aerothermal PIROM is essentially a \CORRECTa{first-order} physics-based component, the turbulent viscous-inviscid interaction (TVI) model \cite{Stollery1969,Stollery1973}, augmented with a data-driven component for enhanced accuracy.  The PIROM formulation is presented in a general manner, so that one can transfer the formulation to other applications \CORRECTa{having, e.g., first-principle physics-based models.}

\subsection{First-order physics-based model}
The TVI model is a classical semi-analytical method, based on a set of \textit{differential-algebraic equations} (DAE's), that predicts the aerothermal load over a deformed and heated slender structure at a given operating condition.  The model is based on the integral momentum equation, obtained from the Navier-Stokes equations through a series of assumptions: \CORRECTa{1) the flow is steady, 2) there is no streamwise heat or momentum transfer, 3) the flow is two-dimensional, 4) the flow outside the boundary layer is inviscid, and 5) the pressure is constant through the boundary layer.} To apply the PIROM method, the TVI model is written in a general form of nonlinear parametric state-space equations,
\begin{subequations}\label{eqn:tvi}
    \begin{align}
        \vM(\vy,\vc;\vtt)\ddf{\vy}{x} &= \vf(\vy,\vc,x;\vu,\vtt) \label{eqn:state space tvi} \\
        \vc &= \vg(\vy,x;\vu,\vtt) \label{eqn:aux var}\\
        \vz &= \vh(\vy,\vc,x;\vu,\vtt) \label{eqn:z}
    \end{align}
\end{subequations}
where the equations are defined over a spatial domain in $x$-coordinates, and five sets of variables are involved,
\begin{compactenum}
    \item State variables $\vy=(\delta^*, M_e, P_e)^T$: the displacement thickness, Mach number, and pressure at the boundary layer edge.
    \item Input variables $\vu=(T_w,y_w)^T$: the thermoelastic response, i.e., the distributions of wall temperature and deformation.
    \item Output variables $\vz=(P_w, \dot{q}_w)^T$: the aerothermal load prediction, i.e., the distributions of wall pressure and heat flux.
    \item System parameters $\vtt=(M_\infty,P_\infty,T_\infty)^T$: the freestream conditions.
    \item Auxiliary variables $\vc=(H,C_f,k_p)^T$: the boundary layer shape factor, skin friction coefficient, and the ratio between $P_w$ and $P_e$.
\end{compactenum}
\CORRECTa{Note that while the TVI model is developed for a steady boundary layer distribution, it can be viewed as a dynamical system with coordinates $x$ that governs the growth of the boundary layer along the streamwise direction.}

In Eq. \eqref{eqn:state space tvi} the system matrix and the state dynamics are, respectively,
\begin{equation}
    \vM(\vy,\vc;\vtt) = \left(\begin{matrix}
        H   & \kappa\frac{H\delta^*}{M_e}-2\delta^*M_e\xi & 0 \\
        M_e & 0 & 0 \\
        0   & 0 & 0
    \end{matrix}\right),\quad
    \vf(\vy,\vc,x;\vu,\vtt) = \left(\begin{matrix}
        H^2\frac{C_f}{2} \\
        \alpha - M_e\frac{dy_w}{dx} \\
        P_e - P_o\left(1 + \frac{\gamma-1}{2}M_e^2\right)^{\frac{-\gamma}{\gamma-1}}
    \end{matrix}\right)
\end{equation}
where \CORRECTa{the following factors are introduced for convenience},
\begin{equation}
    \alpha = \frac{2}{\left(\gamma-1\right)}\left(\left(\frac{P_e}{P_\infty}\right)^{\frac{\gamma-1}{2\gamma}} - 1\right),\quad
    \xi = \frac{\gamma-1}{2}\left(1+H_i\frac{T_w}{T_o}\right),\quad
    \kappa = H_i\frac{T_w}{T_o}-4
\end{equation}
\CORRECTa{The $3\times 3$ system matrix $\vM$ in Eq. (\ref{eqn:state space tvi}) is at most rank-2 and makes the TVI model a DAE system.}

\CORRECTa{Next, Eq. (\ref{eqn:aux var}) correlates the auxiliary variables to the other variables. The skin-friction coefficient $C_f$ is obtained from Eckert's reference enthalpy method \cite{Brouwer2018}. The shape factor $H$ is modeled using Crocco's \cite{Stollery1969,Stollery1973} boundary layer shape factor approximation with $H_i$ as the incompressible boundary layer shape factor, which relates the displacement thickness to the momentum thickness $H=\delta^*/\theta$ at any point along the streamwise direction}. The auxiliary variables now become,
\begin{equation}\label{eqn:auxiliary vector1}
    \left(\begin{matrix}
        H \\ C_f \\ k_p
    \end{matrix}\right)
    = \vg(\vy,x;\vu,\vtt) =
    \left(\begin{matrix}
        \frac{\gamma-1}{2}M_e(x)^2\left(1 + H_i\frac{T_w}{T_o}\right) \\
        0.026\frac{T_e}{T_r}\left(\frac{\rho_\infty \mu_r}{\rho_e \mu_e}\right)^{1/4} Re_\theta^{-1/4} \\
        1
    \end{matrix}\right)
\end{equation}
where \CORRECTa{$T_r$ is the reference temperature, and $k_p=1$ is due to the constant pressure boundary layer assumption.}

Finally, Eq. \eqref{eqn:z} extracts the output of interest, i.e., the aerothermal load, from the system states,
\begin{equation}\label{eqn:z output}
    \left(\begin{matrix}
        P_w \\ \dot{q}_w
    \end{matrix}\right)
    = \vh(\vy,\vc,x;\vu,\vtt) = \left(\begin{matrix}
        k_pP_e \\
        \frac{C_f}{2Pr^{2/3}}\rho_e u_e\left(h_{aw}(T_r) - h(T_w)\right)
    \end{matrix}\right)
\end{equation}
\CORRECTa{where the adiabatic wall enthalpy $h_{aw}(T)$ is computed using Eckert's reference enthalpy method \cite{Eckert1956,Eckert1960,Huang2019b}}.

The key feature of the TVI model in terms of the aerothermal load modeling is its \textit{generalizability}.  The inputs to the TVI model, i.e., the distributions of wall temperature and deformation, are arbitrary continuous functions\CORRECTa{, as long as the wavelength of structural deformation is several orders of magnitude larger than the thickness of the boundary layer. Hence the model can be used to predict the aerothermal loads over slender structures with arbitrary continuous thermoelastic responses that satisfy the requirements above.} The generalizability gives a clear advantage to the TVI model over the conventional aerothermal surrogates, which typically require a parametrization of the thermoelastic responses using a finite set of basis functions for a fixed geometrical configuration.

\subsection{Data-driven augmentation}

The main disadvantage of the classical TVI model is its lack of quantitative accuracy in the aerothermal load prediction, when compared to high-fidelity CFD solutions.  The TVI model in the nonlinear parametric state-space form in Eq. (\ref{eqn:tvi}) reflects a common scenario in many engineering applications.  The system dynamics Eq. (\ref{eqn:state space tvi}) and the output equation Eq. (\ref{eqn:z}) are \CORRECTa{typically derived to capture the first-order physics with higher-order effects ignored}, while the auxiliary variable equation Eq. \eqref{eqn:aux var} involves expressions that are based on either semi-empirical correlations or simplifying assumptions, which causes inaccuracies in the predictions.  In the PIROM formulation, augmentation terms for the auxiliary variables are introduced to reduce or eliminate the sources of error due to the model assumptions. The resulting aerothermal model is termed the augmented TVI (ATVI) equations.



\subsubsection{Algebraic Augmentation}

In the classical FIML approach, the augmentation is accomplished in an algebraic manner.  A vector of augmentation functionals $\bbeta = \left(\beta_H, \beta_C, \beta_W\right)^T$, with unknown field functions $\beta_H$, $\beta_C$, and $\beta_W$, are introduced to correct the auxiliary variables, i.e., the boundary layer shape factor $H$, skin friction coefficient $C_f$ and constant pressure factor $k_p$.  The augmentation vector is assumed to be an algebraic equation, e.g., $\vtb=\vtb(\vy,x;\vu,\vtt)$, and the auxiliary equation is modified as $\vc=\tilde{\vg}(\vy,x;\vu,\vtt,\vtb)$, which may take various functional forms, such as,
\begin{compactenum}
\item Additive form: $\tilde{\vg}=\vg+\vtb$, where $\vtb=\vtb(\vy,x;\vu,\vtt)$; basic TVI model is recovered when $\vtb=0$.
\item Multiplicative form: $\tilde{\vg}=\vtb\odot\vg$, where $\vtb=\vtb(\vy,x;\vu,\vtt)$ and $\odot$ is the element-wise product; basic TVI model is recovered when $\vtb=1$.
\item Compositional form: $\tilde{\vg}=\vtb\circ\vg$, where $\vtb=\vtb(\vg;\vy,x;\vu,\vtt)$ and $\circ$ represents functional composition; basic TVI model is recovered when $\vtb$ is an identity mapping.
\end{compactenum}
In this study, the multiplicative form is found to be sufficient for the algebraic augmentation of the TVI model, and the ATVI equations are,
\begin{subequations}\label{eqn:algebraic atvi}
    \begin{align}
        \vM(\vy,\tilde{\vc};\vtt)\ddf{\vy}{x} &= \vf(\vy,\tilde{\vc},x;\vu,\vtt) \\
        \tilde{\vc} &= \tilde{\vg}(\vy,x;\vu,\vtt,\bbeta) \equiv \vtb(\vy,x;\vu,\vtt)\odot\vg(\vy,x;\vu,\vtt) \\
        \vz &= \vh(\vy,\tilde{\vc},x;\vu,\vtt)
    \end{align}
\end{subequations}
where $\vf$, $\vg$ and $\vh$ are the same as in the classical TVI model, Eq. \eqref{eqn:tvi}, while the auxiliary variables $\tilde{\vc}$ are augmented by $\vtb$.  The goal of the augmentation is to make the outputs of the ATVI equation, i.e., the pressure and heat flux distributions, match well with high-fidelity solutions but with significantly lower computational cost.  The unknown function $\vtb$ for the augmentation terms is a data-driven model such as a neural network (NN) or a Gaussian process (GP).  The learning algorithms for the algebraic augmentation terms has been developed in the classical FIML approaches.

\subsubsection{Differential Augmentation}

The algebraic augmentation formulation has found its success in classical FIML applications, especially the turbulence closure modeling for RANS \cite{Duraisamy2013,Duraisamy2016,Holland2019,Holland2019a,Holland2019b}.  However, in these applications, the governing equations typically does not involve time evolution. \CORRECTa{Since the TVI model Eq. (\ref{eqn:tvi}) yield a dynamical system for the growth of the boundary layer along the streamwise direction,} it may be beneficial to assume a dynamical form for the augmentation variables as well to achieve higher predictive accuracy.  This means using a differential augmentation equation, instead of an algebraic one, to determine the $\vtb$ distribution.
The ATVI equations with differential augmentation are,
\begin{subequations}\label{eqn:differential atvi}
    \begin{align}\label{eqn_dat1}
        \vM(\vy,\tilde{\vc};\vtt)\ddf{\vy}{x} &= \vf(\vy,\tilde{\vc},x;\vu,\vtt) \\\label{eqn_dat2}
        \vA\ddf{\vtb}{x} &= \vtG(\vy,\bbeta,x;\vu,\vtt) \\\label{eqn_dat3}
        \tilde{\vc} &= \tilde{\vg}(\vy,x;\vu,\vtt,\bbeta) \\\label{eqn_dat4}
        \vz &= \vh(\vy,\tilde{\vc},x;\vu,\vtt)
    \end{align}
\end{subequations}
where $\vA$ is the unknown augmented system matrix and $\vtG$ is the unknown augmented dynamics.  The matrix $\vA$ can be rank-deficient and the differential augmentation Eq. (\ref{eqn_dat2}) itself can be a DAE; furthermore, when $\vA=\vO$ Eq. (\ref{eqn_dat2}) effectively reduces to an algebraic augmentation.
\CORRECTa{When $\vA$ is full-rank, Eq. (\ref{eqn_dat2}) is effectively the neural ordinary differential equation \cite{Chen2018}; it is sufficient to choose $\vA$ to be an identity matrix and only learn $\vtG$ from data.}
Similar to the case of algebraic ATVI model, $\vtG$ can be represented by a data-driven model such as a NN or a GPR.  However, the learning of the differential augmentation requires a new learning algorithm beyond the classical FIML approach, which will be presented in the following section.


\section{Learning Algorithms for Physics-Infused Reduced Order Modeling}\label{sec3:train}

The PIROM methodology is closely related to the FIML paradigm but further extends it to a more general form that supports both algebraic and differential augmentations.  The main challenge in the development of PIROM's is the determination of the unknown augmentation terms from a dataset of high-fidelity solutions.  Because of the involvement of physics-based equations in PIROM, the conventional learning algorithms for a purely data-driven model, e.g., the stochastic gradient descent algorithms for NN's, cannot be directly applied to learn the data-driven component of the PIROM.  This section presents two learning algorithms for PIROM, including the indirect (PIROM-i) and direct (PIROM-d) approaches.  In the following, it is assumed that the training dataset consists of $N_s$ high-fidelity sample solutions associated with a series of inputs and system parameters, denoted $\{(\vu^i,\vtt^i,\vz^i)\}_{i=1}^{N_s}$.

\begin{figure}
    \centering
    \insertfigs{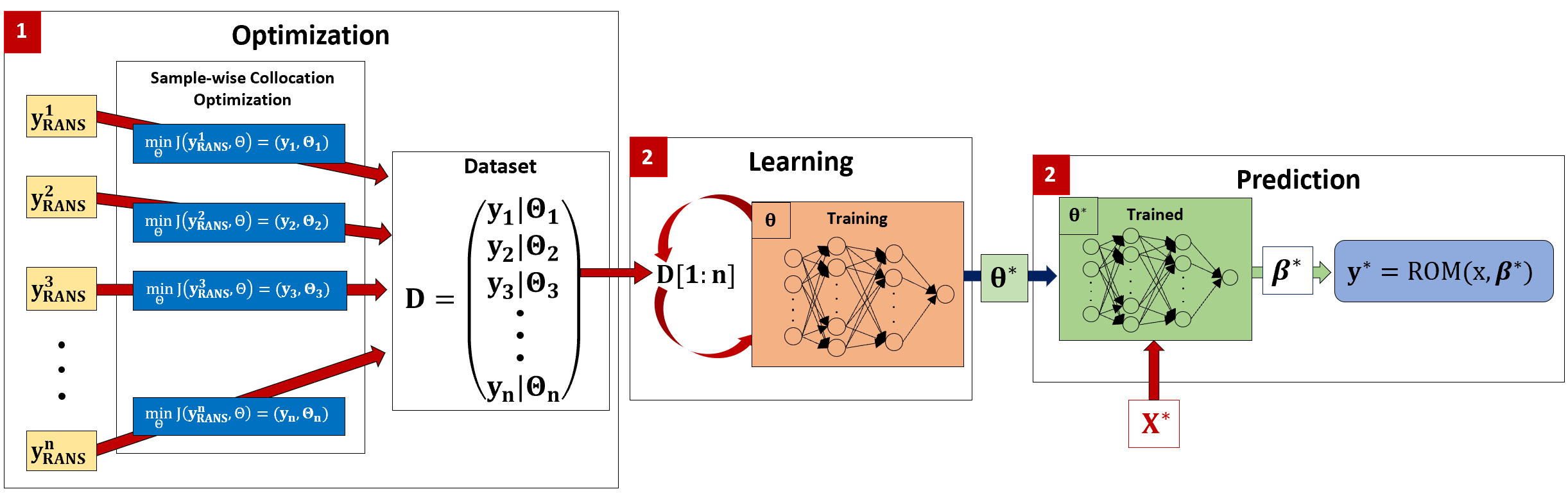}{0.80}{Indirect PIROM approach.}
    \insertfigs{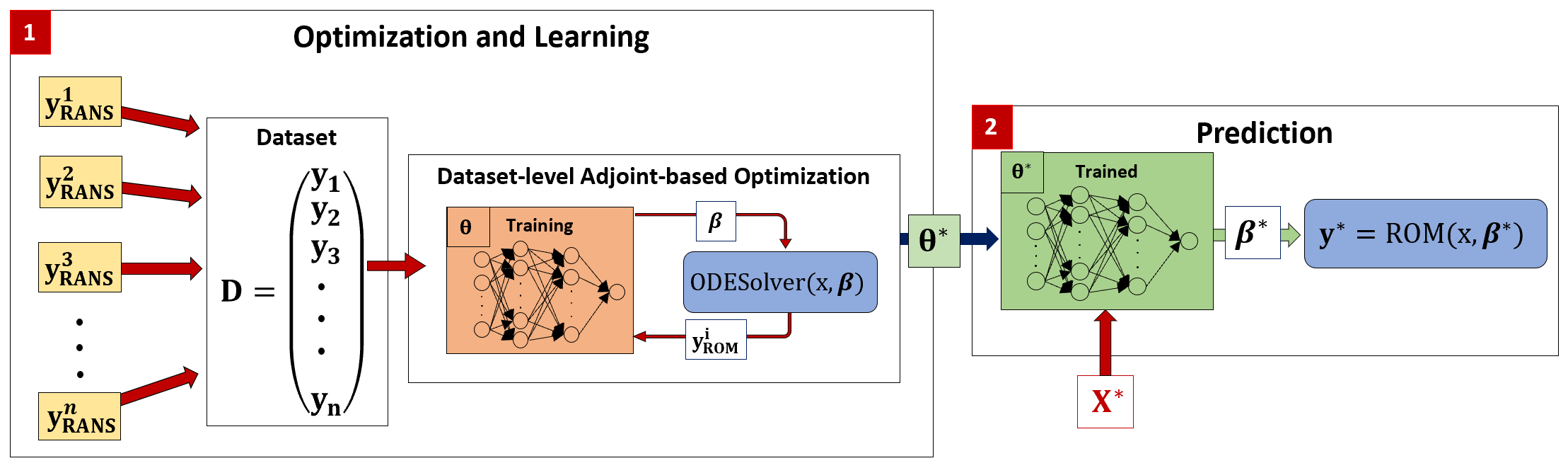}{0.80}{Direct PIROM approach.}
    \caption{Illustration of the PIROM approaches based on FIML indirect and direct techniques.}
    \label{fig:piroms}
\end{figure}

\subsection{Indirect approach for PIROM}

The PIROM-i approach consists of three decoupled stages: optimization, learning, and prediction, as depicted in Fig. \ref{figs/sec3/pirom-i.PNG}.  The first stage performs $N_s$ optimizations, and the $i$th optimization finds the \textit{ad hoc} values of $\vtb$ such that the PIROM can reproduce the high-fidelity solution for the $i$th input $(\vu^i,\vtt^i)$.  The second stage interpolates the $N_s$ sets of \textit{ad hoc} $\vtb$'s to new inputs and produces the desired data-driven component for the PIROM.  The prediction stage involves a specially-designed procedure for enhanced convergence and numerical stability for the ATVI evaluation.  The details are described next.

In the optimization stage, the augmentation variables are represented using a direct collocation approach and parameterized using a discrete set of values along the computational domain,
\begin{equation}
    \vB=(\vtb_{(1)},\vtb_{(2)},\cdots,\vtb_{(M)})
\end{equation}
positioned on $M$ grid points over the computational domain $\{x_{(1)},x_{(2)},\cdots,x_{(M)}\}$. Given the parameters $\vB$, a spline is fitted to represent the augmentation variables and their derivatives in a continuous form,
\begin{equation}\label{eqn:augmentation spline}
    \vtb = \vtb_{spl}(x;\vB),\quad \frac{d\vtb}{dx}=\frac{d\vtb_{spl}}{dx}\left(x;\vB\right)
\end{equation}

Consider the $i$th high-fidelity (HF) solution for the aerothermal loads $\vz^i_{HF}$ that is generated for a thermoelastic response $\vu^i$ and an operating condition $\vtt^i$. A DAE-constrained optimization problem is formulated to find the optimal augmentation values $\vB^i$ such that the ATVI solutions match with $\vz^i_{HF}$ as close as possible,
\begin{subequations}\label{eqn:optimization}
    \begin{align}
        \vB^i &= \arg\min_{\vB} \quad \mathcal{J}(\bz^i_{HF}, \bz^i(\vB)) + \xi\mathcal{J}_\beta(\vB) \label{eqn:opt1} \\
        s.t. \quad \vM\left(\vy,\tilde{\vc};\vtt^i\right)\ddf{\vy}{x} &= \vf\left(\vy,\tilde{\vc},x;\vu^i,\vtt^i\right) \label{eqn:opt2} \\
        \tilde{\vc} &= \vtb_{spl}(x;\vB)\odot\vg(\vy,x;\vu^i,\vtt^i) \label{eqn:opt3} \\
        \vz^i &= \vh(\vy,\tilde{\vc},x;\vu^i,\vtt^i)
    \end{align}
\end{subequations}
The objective functions are,
\begin{subequations}
\begin{align}
    \mathcal{J}(\bz^i_{HF}, \bz^i(\vB)) &= \frac{1}{2} \int_0^L l(\bz^i_{HF}, \vB) dx,\quad l(\bz^i_{HF}, \vB) = \norm{\bz^i_{HF}-\bz^i(\vB)}^2\\
    \mathcal{J}_\beta(\vB) &= \norm{\ddf{\vtb_{spl}(\vB)}{x}}^2
\end{align}
\end{subequations}
where $\mathcal{J}_\beta(\vB)$ is a regularization term for reducing the non-physical oscillations caused by the spline fit and $\xi$ is an user-specified factor.  In this study, $\xi=1\times10^{-5}$.  The results from $N_s$ optimizations form a dataset of optimal \textit{values} for $\bbeta$ associated with different flow solutions,
\begin{equation}\label{eqn:dataset}
    \cD = \left\{(\vw^i_{(j)}, \vtb_{(j)}^i)\ |\ i=1,\cdots,N_s,\ j=1,\cdots,M \right\}
\end{equation}
where $\vw^i_{(j)}=(\vy^i_{(j)},x_{(j)},\vu^i_{(j)},\vtt^i)$ is a compact notation for the variables relevant to the $\vtb$ function.  Note that $\vy^i$ is the ATVI solution obtained with the optimal augmentation variable $\vB^i$ and is expected to produce the desired output $\vz_{HF}^i$.

Subsequently, the second stage of PIROM-i consists of training a data-driven model to accurately represent the augmentation $\bbeta$ as a function of the inputs $\vw$, so that the correct augmentation values can be obtained for the inputs that are different from those in the training dataset.  In this study, the standard GP model is employed \cite{Rasmussen2006},
\begin{equation}\label{eqn:gpr augmentation}
    \vtb = \vtb_{GP}(\vw;\cD)\equiv \vtb_{GP}(\vy,x;\vu,\vtt;\cD)
\end{equation}

In the third and final stage of PIROM-i, the learned model is incorporated into the ATVI equations as in Eqs. \eqref{eqn:algebraic atvi}. To improve the accuracy and numerical robustness of the ATVI solution, a fixed-point iteration (FPI) algorithm is developed. The FPI algorithm defines two subsystems from Eq. \eqref{eqn:algebraic atvi} and \eqref{eqn:gpr augmentation},
\begin{subequations}\label{eqn:fpi}
    \begin{align}
        &\left\{\begin{array}{l}
            \vM(\vy,\tilde{\vc};\vtt)\ddf{\vy}{x} =     \vf(\vy,\tilde{\vc},x;\vu,\vtt) \\
            \tilde{\vc} = \vtb^* \odot \vg(\vy,x;\vu,\vtt) \\
            \vz = \vh(\vy,\tilde{\vc},x;\vu,\vtt)
        \end{array}
        \right.\label{eqn:subsystem1} \\
        &\left\{\vtb^* = \vtb_{GP}(\vy,x;\vu,\vtt;\cD)
        \right.\label{eqn:subsystem2}
    \end{align}
\end{subequations}

The FPI algorithm begins with an initial guess $\vtb_{(0)}=1$ for the augmentation variables.  Starting from $k=1$, at the $k$th iteration a new flow solution $\vy_{(k)}$ is solved using Eq. \eqref{eqn:subsystem1} with $\vtb_{(k-1)}$.  Then, a new augmentation variable $\vtb_{(k)}$ is computed using Eq. (\ref{eqn:subsystem2}) with $\vy_{(k)}$.  The sequence of $(\vy_{(k)},\vtb_{(k)})$ is generated until convergence. 
To accelerate the convergence of the algorithm, an auxiliary GP model $\bbeta_{AG}$ is trained with the following dataset,
\begin{equation}\label{eqn:dataset}
    \hat{\cD} = \left\{(\hat{\vw}^i_{(j)}, \vtb_{(j)}^i)\ |\ i=1,\cdots,N_s,\ j=1,\cdots,M \right\}
\end{equation}
where $\hat{\vw}^i_{(j)}=(\hat{\vy}^i_{(j)},x_{(j)},\vu^i_{(j)},\vtt^i)$ and $\hat{\vy}^i$ is the solution of classical TVI model given the input $\vu^i$ and system parameters $\vtt^i$.
The model $\vtb_{AG}$ provides an improved initial guess of augmentation variables for the FPI algorithm. The complete FPI algorithm is summarized in Alg. \ref{alg:fpi}, where $N_{itr}$ and $\epsilon$ are the user defined maximum number of iterations and error threshold, respectively. Numerical experiments indicate that $N_{itr}\approx5$ iterations are sufficient with a threshold of $\epsilon\approx0.01$.

\begin{algorithm}
    \SetAlgoLined
    Given thermoelastic response $\vu$ and flow conditions $\vtt$. \\
    Solve Eq. (\ref{eqn:subsystem1}) with $\vtb_{(0)} = \boldsymbol{1}$ to obtain the classical TVI solutions $\hat{\vy}$. \\
    Compute the initial guess of augmentations variables using the auxiliary GP: $\vtb^*_{(1)}=\vtb_{AG}(\hat{\vy},x;\vu,\vtt;\hat{\cD})$. \\
    Set $k=1$ and tolerance $\epsilon$. \\
    \While{$k \leq N_{itr}$ and $e\geq\epsilon$}{
    Solve Eq. (\ref{eqn:subsystem1}) using $\bbeta_{(k)}$ to obtain state variables $\vy_{(k)}$ and output variables $\vz_{(k)}$. \\
    Update the augmentation values $\vtb_{(k+1)}=\vtb_{GP}(\vy_{(k)},x;\vu,\vtt;\cD)$. \\
    Compute the error $e=\norm{\vtb_{(k+1)}-\vtb_{(k)}}$. \\
    $k=k+1$.}
    Return the final output $\vz_{(k)}$ as the predicted steady aerothermal load.
    \caption{Fixed-point iteration algorithm for PIROM-i.}
    \label{alg:fpi}
\end{algorithm}

The PIROM-i is relatively easy to implement and preliminary studies have successfully utilized the PIROM-i for aerothermal load modeling applications \cite{Venegas2021,Venegas2022}.  However, the main disadvantages of PIROM-i are two-fold.  One is that solving the $N_s$ optimization problems can be time consuming.  The other is that the optimization stage may produce a dataset that is not entirely learnable by the data-driven model in the subsequent learning stage. The set of flow features that are related to the optimal augmentation variables $\vtb$ may not contain enough correlation for the data-driven model to accurately represent the augmentation function.


\subsection{Direct approach for PIROM}

The direct PIROM (PIROM-d) resolves the limitations of PIROM-i by directly embedding the learning stage into the optimization stage as shown in Fig. \ref{figs/sec3/pirom-d.PNG}, which has already been done in classical FIML methods for algebraic augmentations.  This work generalizes the training algorithm to the differential augmentation case.


First, unlike the PIROM-i approach, only one optimization problem is solved for PIROM-d,
\begin{subequations}
    \begin{align}\label{eqn:direct nn2}
        \vtQ^* &= \arg\min_{\vtQ} \quad \sum_{i=1}^{N_s} \mathcal{J}(\bz^i_{HF}, \bz^i(\vtQ)) \\\label{eqn_pd_c1}
        s.t. \quad \vM(\vy,\tilde{\vc};\vtt^i)\ddf{\vy}{x} &= \vf(\vy,\tilde{\vc},x;\vu^i,\vtt^i) \\\label{eqn_pd_c2}
        \vA\ddf{\bbeta}{x} &= \vtG(\vy,\bbeta,x;\vu^i,\vtt^i;\vW) \\
        \tilde{\vc} &= \tilde{\vg}(\vy,x;\vu^i,\vtt^i,\bbeta) \label{eqn_pd_c3} \\ \label{eqn_pd_c4}
        \vz^i &= \vh(\vy,\tilde{\vc},x;\vu^i,\vtt^i),\quad \mbox{for }i=1,\cdots,N_s
    \end{align}
\end{subequations}
where the unknown dynamics $\vtG$ is assumed to be a neural network, parametrized by $\vW$ and the learnable parameters include $\vtQ=\left\{\vA,\vW\right\}$.

Due to the high-dimensionality of the parameters in a typical neural network, the only feasible approach to train the nonlinear differential data-driven component is through gradient-based methods, which requires the sensitivity of the objective function $\mathcal{J}$ with respect to $\vtQ$. Obtaining the gradients $\nabla_{\bTheta}\mathcal{J}$ requires the so-called reverse-mode differentiation through the differential equation solver \cite{Chen2018,Venegas2022}, which boils down to the solution of the adjoint equation of the ATVI model, as presented below.

To avoid the treatment of rank-deficient Jacobians of DAE's that may cause numerical issues, the DAE-constrained optimization problem is converted to an ODE-constrained one by eliminating all the algebraic constraints.  Specifically, the algebraic variables, such as $\tilde{\vc}$ and $P_e$ in $\vy$, are expressed as closed-form functions in terms of the differential variables, such as $(\delta^*,M_e)$ in $\vy$, and the other variables $\vu$ and $\vtt$.  This transforms Eqs. (\ref{eqn_pd_c1})-(\ref{eqn_pd_c3}) to an ODE system, formally written as,
\begin{equation}\label{eqn:implicit constraints}
    \boldsymbol{0} = \vF\left(\bar{\vy},\dot{\bar{\vy}},x;\vu,\vtt;\bTheta\right)
\end{equation}
where $\dot{\Box}=\frac{d\Box}{dx}$ and $\bar{\vy} = \left(\vy,\bbeta\right)^T$ is the augmented state vector.  Furthermore, the remaining algebraic constraint Eq. (\ref{eqn_pd_c4}) can be removed by explicitly incorporating it into the objective function.

Next, without loss of generality, consider just one high-fidelity solution.  The adjoint sensitivity method consists of introducing a Langrange multiplier, i.e., the adjoint variables, $\blambda=\left(\blambda_{\vy}(x),\blambda_{\bbeta}(x)\right)^T$ to remove the constraints from the optimization problem,
\begin{equation}\label{eqn:lagrangian}
    \mathcal{L}(\vz_{HF},\vz\left(\vtQ\right),\blambda) = \int^L_0\left(l(\vz(\vtQ),\vz_{HF}) + \blambda^T\vF\left(\barvy,\dotbarvy,x;\vu,\vtt,\vtQ\right)\right)dx
\end{equation}
Through a variational approach, an extremum of the Lagrangian function in Eq. \eqref{eqn:lagrangian} is obtained when $\delta\mathcal{L}=0$. The following set of conditions are necessary for the parameters $\vtQ^*$ to provide an extremum of the Lagrangian function \cite{Bryson1969,Lewis2012},
\begin{subequations}
\begin{align}
    \vF\left(\barvy,\dotbarvy,x;\vu,\vtt,\vtQ\right) &= \boldsymbol{0} \label{eqn: dynamical constraints}\\
    \left.\left(\ppf{\vF}{\dotbarvy}\blambda\right)^T\delta\barvy\right|_0 &= 0 \label{eqn:left lagrange bc} \\ 
    \left.\left(\ppf{\vF}{\dotbarvy}\blambda\right)^T\delta\barvy\right|_L &= 0 \label{eqn:right lagrange bc} \\
    \ppf{\vz}{\barvy}^T\ppf{l}{\vz} + \ppf{\vF}{\barvy}\blambda - \frac{d}{dx}\left(\ppf{\vF}{\dotbarvy}\blambda\right) &= \boldsymbol{0} \label{eqn:adjoint} \\
    \ppf{l}{\vtQ} + \ppf{\vF}{\vtQ}\blambda &= \boldsymbol{0} \label{eqn:stationarity}
\end{align}
\end{subequations}
Collectively, Eqs. \eqref{eqn: dynamical constraints}-\eqref{eqn:right lagrange bc} are known as the \textit{necessary conditions for an optimum} \cite{Bryson1969}. The condition in Eq. \eqref{eqn: dynamical constraints} reproduces the dynamical constraint in Eq. \eqref{eqn:implicit constraints}. The condition in Eq. \eqref{eqn:left lagrange bc} is automatically satisfied since the initial condition $\barvy(0)$ is known exactly, thus $\delta\barvy(0) = 0$. Since $\barvy(L)$ is arbitrary, Eq. \eqref{eqn:right lagrange bc} provides the value of the Lagrange multiplier at the final location $\blambda(x=L)$. The Eq. \eqref{eqn:adjoint} is referred to as the \textit{adjoint} equation, and it provides the distribution for the Lagrange multipliers by solving the adjoint equation backwards in space using the initial condition from Eq. \eqref{eqn:right lagrange bc}. At $x=0$, $\barvy$ is known but $\blambda$ is unknown. At $x=L$, $\barvy$ is unknown but $\blambda$ is known. Hence the solution of the dynamical system in Eq. \eqref{eqn: dynamical constraints} and the adjoint Eq. \eqref{eqn:adjoint} results in a \textit{two-point boundary value problem} (TPVBP) \cite{Lewis2012,Bryson1969}. Finally, the condition in Eq. \eqref{eqn:stationarity} is known as the \textit{stationarity} condition and ensures that, given the values of the Lagrange multipliers, the parameters $\vtQ^*$ yield an extremum of the Lagrangian function.

For computer implementation, the \texttt{torchdiffeq} Python package \cite{Chen2018,Chen2021} is used to perform reverse-mode differentiation through the ODE solver to determine the gradients using the adjoint sensitivity method \cite{Pontryagin1962} with the help of automatic differentiation \cite{Paszke2017}.

\section{Verification of the PIROM for Aerothermal Modeling}\label{sec:verification}

In this section, the PIROM-i and PIROM-d methods are used to build ROM's for hypersonic aerothermal load prediction over a semi-infinite panel configuration, which is commonly used as a model problem in hypersonic aerothermoelastic studies \cite{Crowell2010,Crowell2011,Crowell2012}.  The PIROM-i method produces an ATVI model with algebraic augmentation, while the PIROM-d method produces an ATVI model with differential augmentation.  The PIROM-i and PIROM-d models are benchmarked against a conventional aerothermal surrogate, i.e., the POD-kriging model, to assess the predictive accuracy and generalization capability.


\subsection{Geometric Configuration}

The aerothermal ROMs are developed for the two-dimensional panel configuration shown in Fig. \ref{figs/sec4/domain2.PNG}, where the CFD mesh and flowfield are depicted for an example thermoelastic response and operating condition. The computational domain is split into the front, panel, and rear sections with lengths $L_f=1$ m, $L_p=1$ m, and $L_r=1$ m, respectively, with a height of $L_h=0.5$ m and panel thickness of $h=0.005$ m. The walls in the front and rear sections are fixed, whereas the panel section is compliant and heat-conducting. The structured fluid mesh consists of 203 points in the x-direction with 73 over the panel section, and 105 normal to the surface. The wall-normal spacing is $\Delta z = 5.0\times10^{-6}$ for sufficient resolution of the boundary layer. All CFD-based simulations are conducted using the extended HYPersonic AeroThermoeElastic (HYPATE-X) computational framework \cite{Huang2018b,Huang2020b}.

\insertfig{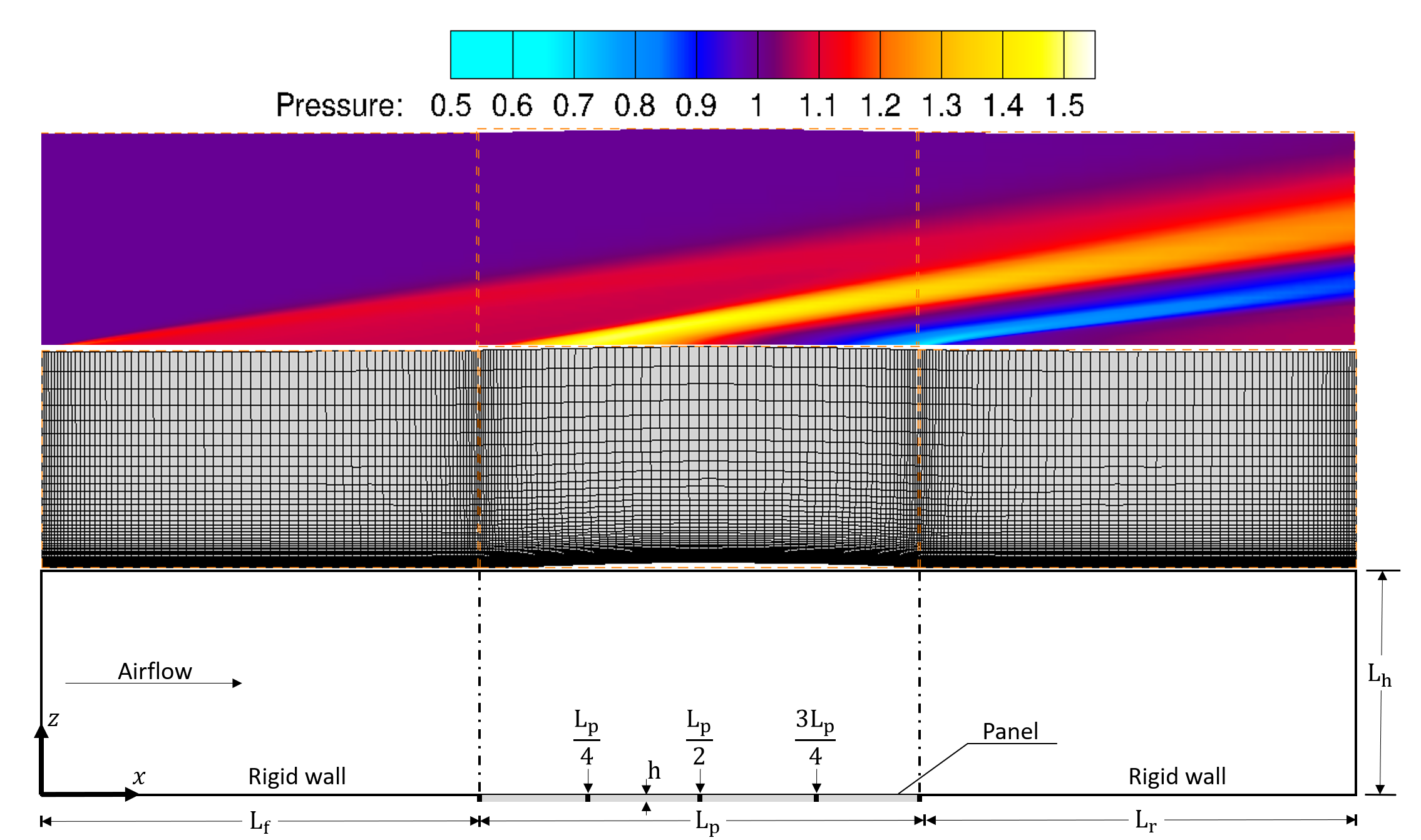}{0.80}{Computational domain of the two-dimensional skin panel configuration.}

\subsection{Aerothermal Models}\label{subsec:aerothermal models}

The implementation details of the three aerothermal ROM's, i.e., PIROM-i, PIROM-d, and POD-kriging, are provided below.

\subsubsection{PIROM}\label{subsubsec:PIROM implementation details}
In PIROM-i, each DAE-constrained optimization problem in Eq. \eqref{eqn:optimization} is solved using the BFGS algorithm \cite{Nocedal1989}, where the gradients are provided by the finite-difference method. In the learning stage, GP models with anisotropic Matern-52 kernels are fitted to represent the augmentation functions $\vtb$, as in Eq. \eqref{eqn:gpr augmentation}. In addition, auxiliary GP models with anisotropic Matern-52 kernels are trained for the convergence acceleration in the prediction stage. All the GP models are implemented using the Python package \texttt{scikit.learn} \cite{Pedregosa2011}.


In PIROM-d, three NN's are used to represent the dynamics of the augmentation variables $\vtb=\left(\beta_H,\beta_C,\beta_W\right)^T$, respectively. \CORRECTa{A preliminary trade study shows that ODE's are sufficient to capture the dynamics of $\vtb$, hence the augmentation system matrix $\vA$ is set to be an identity matrix.}
To increase the learning capabilities of PIROM-d, a \textit{partitioned} NODE formulation is adopted,
\begin{equation}
\ddf{\vtb}{x} = \vtG(\vy,\bbeta,x;\vu,\vtt;\vW) = \vtG_0(\vy,\vtb,x;[0,T_w],\vtt;\vW_0) + y_w(x)\vtG_1(\vy,\vtb,x;[y_w,T_w],\vtt;\vW_1)
\end{equation}
where $\vtG_0$ and $\vtG_1$ are two fully-connected neural networks with weights $\vW_0$ and $\vW_1$, respectively. In the partitioned NODE, the first NN $\vtG_0$ captures the dynamics of boundary layer growth for the flat plate case, i.e. when $y_w(x) = 0$, with possibly non-uniform wall temperature.  The second NN $\vtG_1$ is activated only when $y_w(x)\neq 0$ and accounts for the deviation of boundary layer growth from flat plate case when structural deformation is present. After a series of convergence studies, the NN's describing the $\beta_H$ and $\beta_C$ dynamics consists of ten hidden layers for $\vtG_0$, and eight hidden layers for $\vtG_1$. The NN for the $\beta_W$ dynamics consists of seven hidden layers for both $\vtG_0$ and $\vtG_1$. All the hidden layers employ a ReLU activation function, $f(x)=\max(0,x)$.

\subsubsection{POD-kriging}\label{subsubsec: pod-kriging implementation details}
The aerothermal ROM based on the POD-kriging method maps an input vector $\vd$, representing a parametrized thermoelastic response and an operating condition, to the aerothermal load $\vz$, represented using a set of basis vectors $\vtY$.  The POD-kriging method is well-known \cite{Crowell2011,Falkiewicz2011} and this study adopts the formulation with a physics-informed correction that has been verified for a 2D panel configuration \cite{Huang2019}.  In this formulation, the structural deformation is represented using three sinusoidal modes,
\begin{equation}\label{eqn:kriging elastic parametrization}
    \frac{y_w(x)}{h} = \sum_{i=1}^3\bar{a}_i\sin\left(i\pi\frac{x-L_p}{L_p}\right)
\end{equation}
and wall temperature is represented using the average value $\bar{T}_w$.  The input vector is defined as
\begin{equation}\label{eqn:kriging parmetrization}
    \mathbf{d} = \left[\bar{a}_1,\bar{a}_2,\bar{a}_3,\bar{T}_w\right]
\end{equation}
Ten POD modes $\vtY$ are employed as the basis vector to represent the aerothermal load distribution, with a construction error of less than 1\%.  The complete surrogate model is written formally as
\begin{equation}\label{eqn_pkrg}
    \vz = \vf_{cor}\left(\vtY\vf_{GP}(\vd),T_w(x),\vtt\right)
\end{equation}
where $\vf_{GP}$ is a GP model that maps the input to the POD modal coordinates and $\vtY\vf_{GP}$ approximates the aerothermal load given the sinusoidal deformation and the uniform wall temperature specified by $\vd$.  The anisotropic Matern-52 kernel is used for the GP model.  Subsequently, a pointwise correction $\vf_{cor}$ is employed to account for the effects of non-uniform wall temperature distribution and variations in the operating conditions \cite{Huang2019b}, and produces the final aerothermal load prediction.

\subsection{Datasets for Aerothermal Models}\label{subsec: datasets for aerothermal models}

Next, the generation of the training and test datasets for the aerothermal ROM's are presented. Due to the differences in their formulations, the \CORRECTa{POD-kriging and PIROM models} are learned using two \textit{different} training datasets. \CORRECTa{The aerothermal ROMs are tested on the \textit{same} test dataset.}

\subsubsection{Training Dataset for POD-Kriging}\label{subsubsec: pod-kriging dataset}

The Optimal Latin Hypercube Sampling (OHLS) algorithm is used to populate the four-dimensional parameter space with bounds for each of the dimensions given in Table \ref{table:kriging ranges}. \CORRECTa{The freestream conditions for all the training samples are set to fixed values, and are randomly selected such that these do \textit{not} coincide with any of the freestream conditions in the training dataset for PIROM.} The freestream conditions are set to be $M_\infty=7.523$, $P_\infty=3759.678$ Pa, $T_\infty=466.200$ K. Fixing the freestream conditions also minimizes the potential error introduced by the correction term $\vf_{cor}$ in the POD-kriging formulation Eq. (\ref{eqn_pkrg}).  Based on previous studies \cite{Huang2019,Huang2019b}, 500 training samples are sufficient to achieve convergence with errors less than 3\% in the entire input parameter space.
\CORRECTa{Note that training the POD-kriging model with $N_s$ RANS solutions only counts as $N_s$ samples. Hence, it is typical to encounter sampling requirements of several hundreds to several thousands of high-fidelity flow solutions to obtain a desired level of accuracy from POD-kriging when a higher input dimension is involved.}

\begin{table}
    \caption{Bounds on the POD-kriging input parameter space.}
    \label{table:kriging ranges}
    \begin{center}
        \begin{tabular}{c}
            \hline
            \hline
            $-4.0 \quad \leq \quad \bar{a}_1 \quad \leq 4.0$ \\
            $-2.0 \quad \leq \quad \bar{a}_2 \quad \leq 2.0$ \\
            $-2.0 \quad \leq \quad \bar{a}_3 \quad \leq 2.0$ \\
            $0.92 \quad \leq \quad \bar{T}_w \quad \leq 1.44$ \\
            \hline
            \hline
        \end{tabular}
    \end{center}
\end{table}

\subsubsection{Training Dataset for PIROM Models}\label{subsubsec: PIROM dataset}

The training samples for the PIROM models consist of RANS solutions associated with typical hypersonic thermoelastic responses under different boundary conditions with different Mach numbers $M_\infty$.  Four deformation distributions $y_j(x)$, $j = 1,2,3,4$ and five temperature distributions $T_i(x)$, $i = 1,2,3,4,5$ are selected, as depicted in Fig. \ref{figs/sec4/thermoelastic.png}, where the displacements and the temperature distributions are nondimensionalized by the panel thickness $h=5$ mm and a reference temperature $T_{ref}=380$ K, respectively.  The deformation and temperature distributions are represented using 11th-order polynomials, whose coefficients are provided in the App. \ref{appendix:coefficients}.  Subsequently, the structural and thermal responses are parametrized as follows,
\begin{equation}
    \begin{array}{lll}
        Dj:&\quad y_{w,j}(x,A) = Ay_j(x), &\quad j = 1,2,3,4 \\
        Ti:&\quad T_{w,i}(x,B) = T_{ref} + BT_i(x), &\quad i = 1,2,3,4,5
    \end{array}
\end{equation}
where $A$ is the structural amplitude, and $B$ the temperature amplitude.



\insertfig{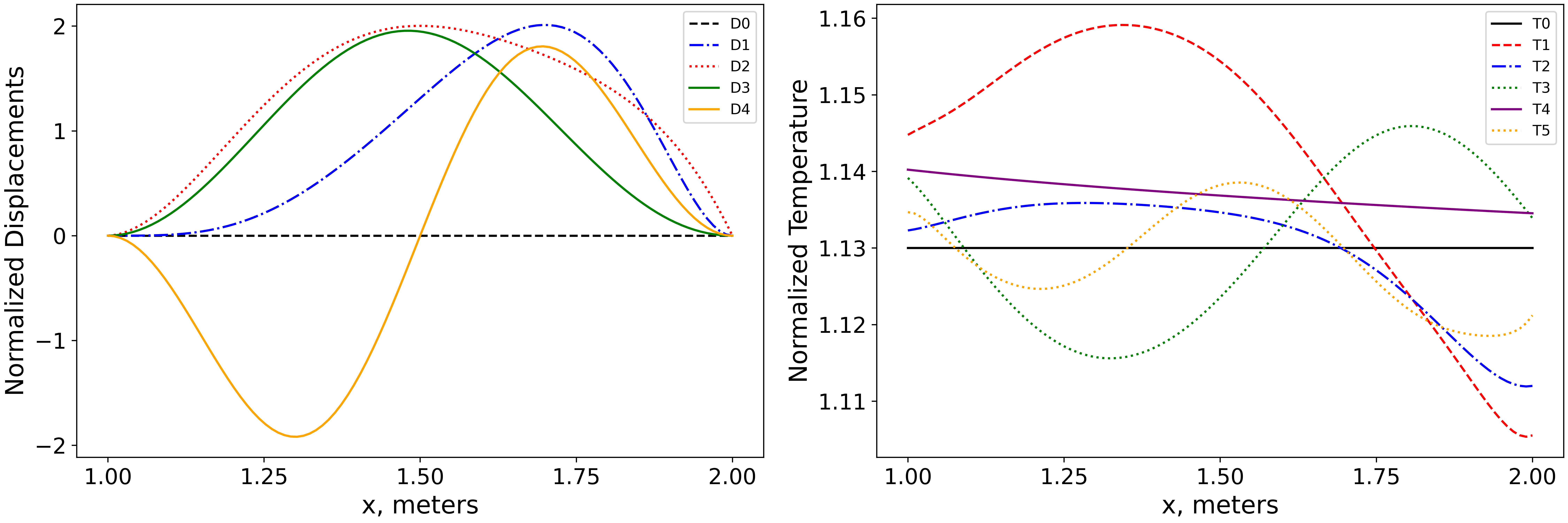}{0.90}{Training thermoelastic modes used in training the PIROMs.}

The chosen structural and thermal responses produce 20 typical thermoelastic responses, each of which is characterized by \CORRECTa{five parameters $(A, B, M_\infty, P_\infty, T_\infty)$}.  For each thermoelastic response, the combinations $(A, B, M_\infty)$ are sampled independently using the OLHS algorithm. \CORRECTa{Then, for each $M_\infty$ the freestream pressure and temperature are uniformly sampled in the intervals $[3580.0, 4200.0]$ Pa and $[340.0,500.0]$ K, respectively, to obtain the freestream conditions $(M_\infty,P_\infty,T_\infty)$ that are uniquely defined for the sample $(A, B, M_\infty)$.  This sampling strategy also ensures the uniform coverage of freestream conditions in the training dataset.} In total, 150 training samples are generated \CORRECTa{in the $(A, B, M_\infty)$ space} over the 20 thermoelastic responses, as depicted in Fig. \ref{figs/sec4/parameter_space.png}. Note that 10 flat-plate samples with $A=0$ and $B=0$ at different Mach numbers, labelled by ``T0'' in Fig. \ref{figs/sec4/parameter_space.png}, are added to enhance the accuracy of the PIROM when the thermoelastic response is small.

\CORRECTa{Note that for each PIROM training sample, the PIROM exploits the high-fidelity information over the panel section in a pointwise fashion. This introduces substantial sampling advantages for PIROM over POD-kriging. In either PIROM-i or PIROM-d, the RANS solutions are not directly incorporated into the training data; instead, the flow variables \textit{at each grid point} in each RANS solution serves to generate one training point. One RANS solution contains $M\approx 100$ grid points, and $N_s\approx 100$ RANS solutions generates $N_s\times M$ samples.  Therefore a few thousands of samples can be obtained from only a dozen of high-fidelity flow solutions for the PIROM.}



\insertfig{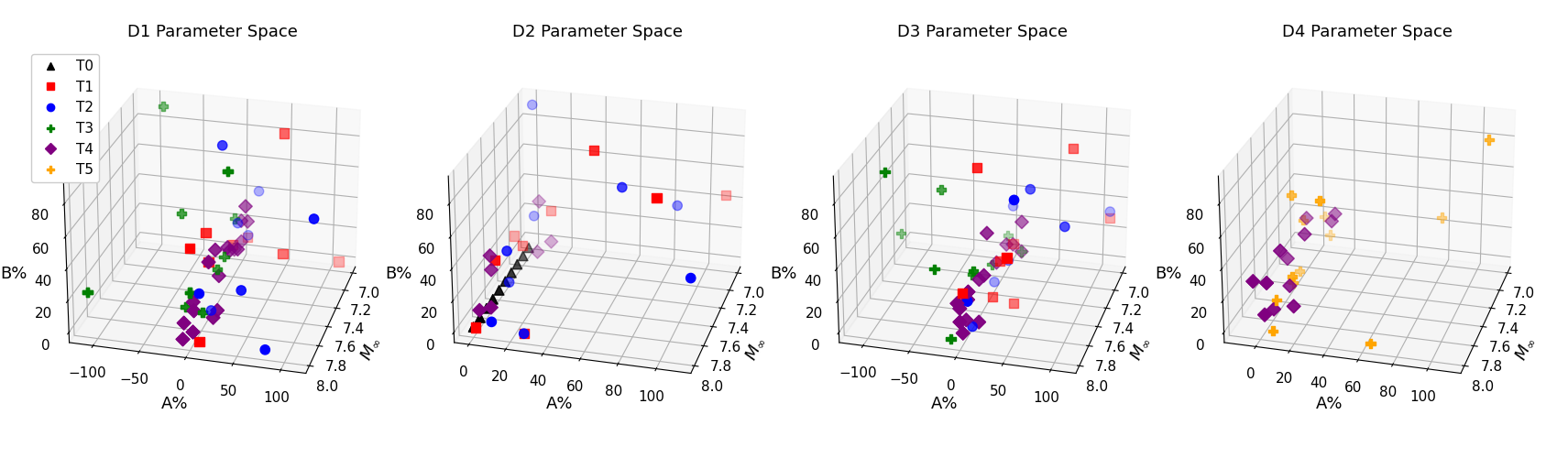}{0.99}{Training samples generated by OLHS and AS techniques.}

\subsubsection{Test Dataset}

The \CORRECTa{test dataset is} defined based on the parametrization used by the PIROM training dataset. All the test samples are generated at the \textit{same} operating conditions as the POD-kriging \CORRECTa{training} dataset, which are different from any operating conditions in the PIROM \CORRECTa{training} dataset. \CORRECTa{The selection of operation conditions gives POD-kriging a modeling advantage over PIROM when the ROMs are used for aerothermal load prediction on the test samples, as the POD-kriging already has information about the freestream conditions. A new thermoelastic response that is not present in any of the POD-kriging or PIROM training datasets is defined for the test samples,}
\begin{equation}
    \begin{array}{ll}
    \text{DY}:&\quad y_w(x,A) = A\left(\dfrac{1}{4}\sum_{j=1}^4 y_j(x)\right) \\
    \text{TY}:&\quad T_w(x,B) = T_{ref} + B\left(\dfrac{1}{5}\sum_{i=1}^{5}T_i(x)\right)
    \end{array}
\end{equation}
where again $A$ and $B$ are the structural and temperature amplitudes, respectively. The \CORRECTa{test} dataset consists of 121 samples generated by sweeping over the parameter combination $A\in[0,100]\%$ and $B\in[0,100]\%$, in steps of $10\%$. The \CORRECTa{test} dataset is used to assess the \CORRECTa{\textit{generalization}} capabilities of PIROM.

\subsection{Comparison of the Aerothermal Models}

The PIROM-i, PIROM-d, and POD-kriging models are benchmarked using the \CORRECTa{test dataset}. The accuracy is characterized using the normalized root-mean squared error (NRMSE),
\begin{equation}\label{eqn:metric}
    NRMSE(\vf_{True},\vf) = \frac{\frac{1}{\sqrt{N}}\norm{\vf_{True} - \vf}}{\max(\vf_{True}) - \min(\vf_{True})}\times 100 \%
\end{equation}
where $\vf$ and $\vf_{True}$ are flow variables of interest from PIROM and RANS solutions, respectively, and $N$ is the number of grid points over the panel domain.

\subsubsection{Comparison of Predictive Performance}

\CORRECTa{The TVI, POD-kriging, and PIROM aerothermal load predictions for the test samples are compared in Figs. \ref{figs/sec4/pressure-sweep.png}, \ref{figs/sec4/heatflux-sweep.png}, and \ref{figs/sec4/piromd_piromi_rom_M7523DYTY-100000-100000.png}. As expected, the TVI model is grossly inaccurate for all test cases, particularly heat flux as shown in Fig. \ref{figs/sec4/piromd_piromi_rom_M7523DYTY-100000-100000.png}. The PIROM models achieve consistently low errors of less than $3\%$ in all cases, while the errors of POD-kriging model vary from $3\%$ to $13\%$.}

\begin{figure}
    \centering
    \insertfigs{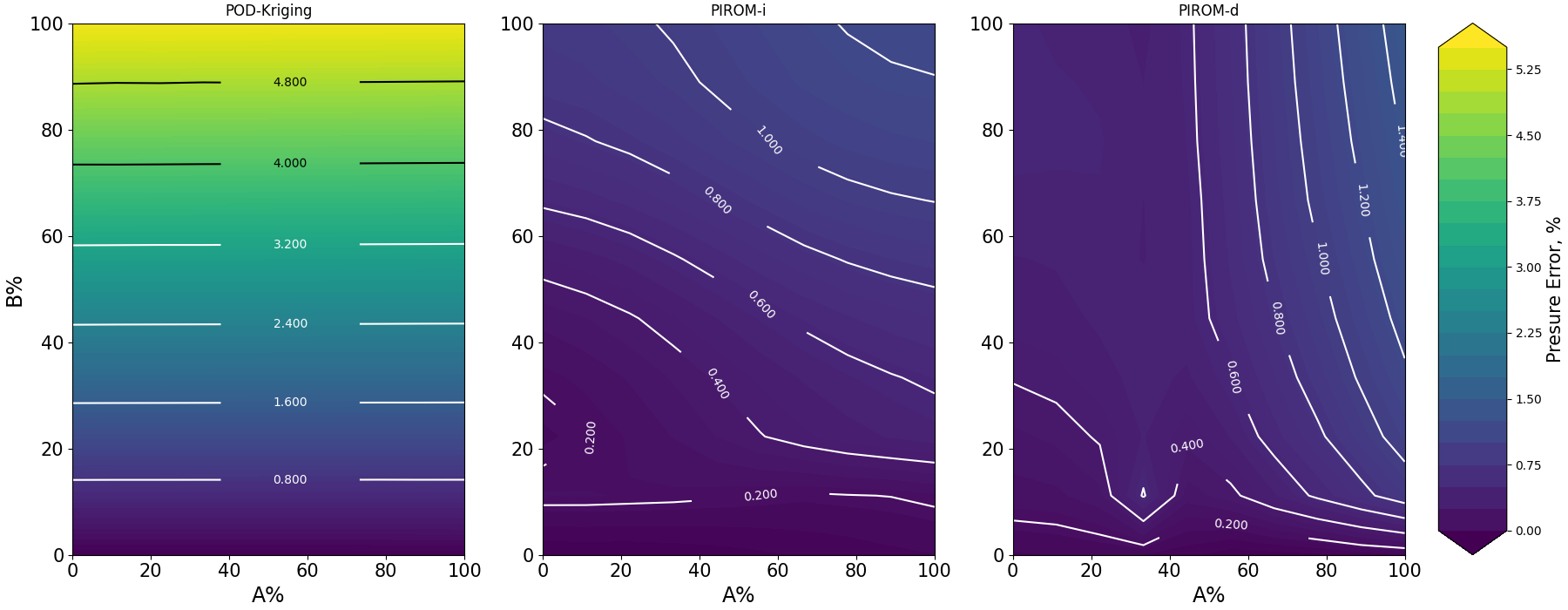}{0.90}{Performance in aerodynamic load prediction.}
    \insertfigs{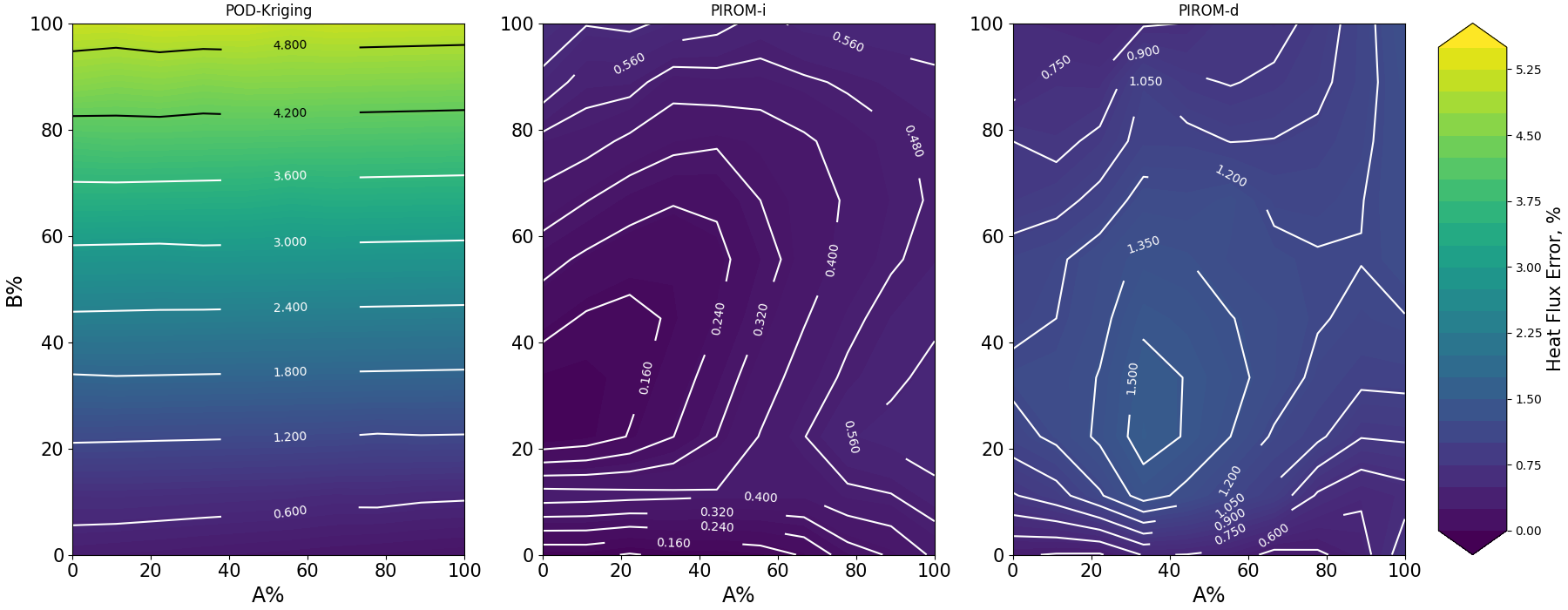}{0.90}{Performance in heat flux load prediction.}
    \insertfigs{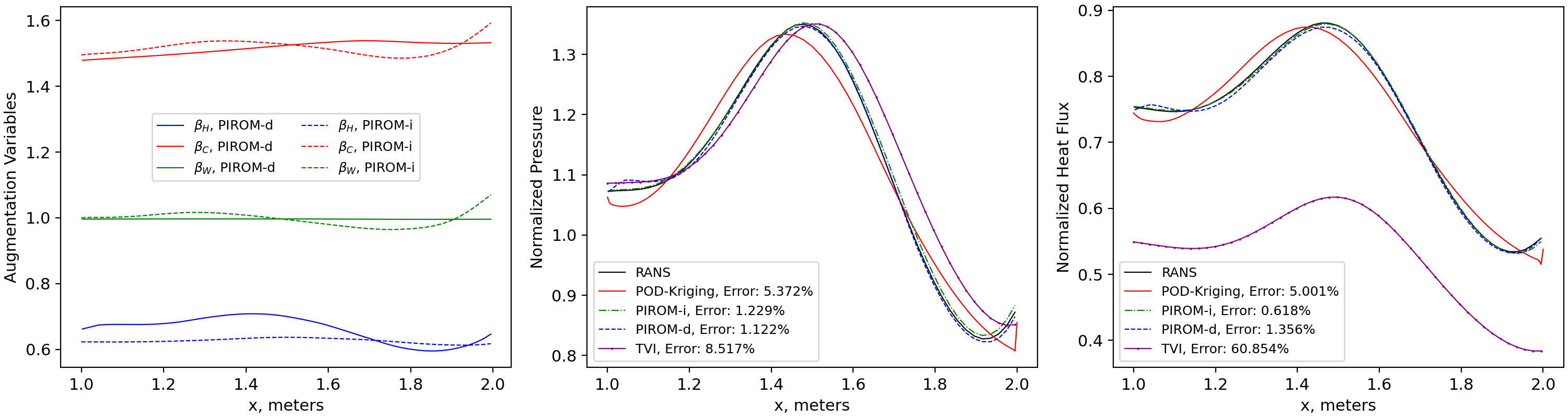}{0.90}{Aerothermal load distribution for $A=100\%$ and $B=100\%$.}
    \caption{Augmentation variables and aerothermal load distributions from PIROM-i, PIROM-d, and POD-kriging for the synthesized thermoelastic response.}
    \label{fig:sweep}
\end{figure}

\CORRECTa{The PIROM and POD-kriging aerothermal load results for the 121 test samples are shown in Figs. \ref{figs/sec4/pressure-sweep.png} and \ref{figs/sec4/heatflux-sweep.png}}. These two aerothermal models accurately predict the pressure and heat flux distributions for thermoelastic responses of low amplitude, and the prediction error tends to increase as the structural and thermal amplitudes increase. This increasing trend of error is attributed to the difficulties in modeling the aerothermal load nonlinearities for structural deformations of large amplitude. However, it is evident from Fig. \ref{fig:sweep} that PIROM-i and PIROM-d outperform POD-kriging on the entire parameter space. Among PIROM-i and PIROM-d, both aerothermal models provide similar accuracy over the parameter space with a slight advantage of PIROM-d in pressure prediction at larger structural and thermal amplitudes.

\CORRECTa{The aerothermal load prediction for the test sample with $A=100\%$ and $B=100\%$ is shown in Fig. \ref{figs/sec4/piromd_piromi_rom_M7523DYTY-100000-100000.png}, where the pressure is normalized by $P_\infty$ and the heat flux is normalized by the heat flux in Eq. (\ref{eqn:z output}) evaluated at freestream quantities. The TVI model yields highly inaccurate pressure and heat flux load predictions over the deforming structure; particularly, the predicted heat flux exhibits large errors in magnitude. The POD-kriging performs with better accuracy when compared to the TVI model, but incorrectly predicts the location and amplitude of the aerothermal load peaks over the deforming structure as shown in Fig. \ref{figs/sec4/piromd_piromi_rom_M7523DYTY-100000-100000.png}. The PIROMs are consistently accurate even if the thermoelastic response is unseen in the training dataset.}

\CORRECTa{Figure \ref{figs/sec4/piromd_piromi_rom_M7523DYTY-100000-100000.png} highlights} the typical characteristics of the PIROM-based aerothermal models.  The augmentation variables \hl{in the ATVI model} not only enable highly accurate aerothermal load predictions \hl{relative to the uncorrected TVI model}, but are also \textit{interpretable}. The $\beta_C$ and $\beta_H$ increase the skin-friction coefficient and reduce the shape factor, respectively. The magnitude of these two augmentation variables are approximately reciprocals of each other, indicating the balance between the forcing terms and auxiliary variables in the ATVI equations. Furthermore, $\beta_C$ increases the heat flux by increasing the skin-friction coefficient, and shifts the peak of the pressure load upstream over the panel section. The $\beta_W$ augmentation variable shows significant deviations from unity over the panel domain, indicating the violation of the constant pressure boundary layer assumptions when shock compression or expansion is present.

Comparing the $\vtb$ distributions in the PIROM-i and PIROM-d models, it is clear that the augmentation variables are non-unique, though the $\vtb$ distributions in the two models follow a similar trend.  The non-uniqueness is attributed to two sources.  One source is the non-convexity of the space of augmentation variables.  This means that there exist different combinations of $\beta_H$, $\beta_C$ and $\beta_W$ functions that provide equivalent information to the ATVI model to achieve a desired level of accuracy.  The other source is the data-driven component of PIROM where the learning of GP or NN is a non-convex optimization that typically converges to a local minimum.  In the computational sense, the $\vtb$ distributions from both PIROM models produce satisfactorily accurate aerothermal predictions, however, it remains future work to determine which distributions are more representative of the actual physical process and how to reliably obtain such distributions in the PIROM training algorithm.

\subsubsection{Remarks on the Comparison}

\CORRECTa{The comparisons on the test samples} show that the PIROM achieves accurate aerothermal prediction, and \CORRECTa{that} such accuracy generalizes well to thermoelastic responses not seen in the training dataset.  In addition, the augmentation variables possess clear physical interpretations that are not available in conventional interpolation-based aerothermal surrogates.

The advantage of PIROM over the conventional method is even more evident if one accounts for the \CORRECTa{differences between the resulting training datasets for each of the aerothermal models. The differences are that (1) the number of CFD solutions for training used for the POD-kriging surrogate is far more than those used in the PIROM, making the PIROM far more computationally tractable,} and (2) the PIROM was trained over a wider range of operating conditions. This means that when testing the aerothermal models at an operating condition for which POD-kriging was specifically trained, the PIROM is clearly at a sampling disadvantage, since it was trained with far less sample points for that specific operating condition.  \CORRECTa{As discussed earlier in Section \ref{subsubsec: PIROM dataset}, the advantage of PIROM in sampling is explained by the fact that PIROM extracts out a multitude of pointwise samples from each CFD solution, producing a much larger and information-rich dataset when compared to the conventional interpolation based methods.}  Thus, besides
the modeling accuracy and generalizability, the PIROM is also far more computationally tractable than the conventional
POD-kriging.


\section{Application to Hypersonic Aerothermoelastic Analysis}\label{sec:results}

The verification of PIROM for aerothermal modeling in the previous section was conducted in an \textit{offline} setting, i.e. the thermoelastic responses were specified \textit{a priori}, and then supplied to the PIROM to generate a steady aerothermal load. This section demonstrates the accuracy and robustness of PIROM in a \textit{online} setting, where the PIROM is utilized as an aerothermodynamic solver and coupled to a thermoelastic solver to conduct fully-coupled transient hypersonic ATE simulations.  The key difference between the offline and online settings is that in the latter case the errors in the ROM may propagate to the thermoelastic solution and the numerical errors may accumulate over time.

\subsection{Definition of Panel Configurations}\label{sec:definition of panel configurations}

The panel configuration considered in the previous section is extended with extra structural constraints to resemble more realistic hypersonic structures where the skin panel is stiffened with ribs.  At the leading and trailing edges, clamped (C), spring-constrained (X), and simply-supported (S) boundary conditions are considered.  Along the panel, additional simply-supported constraints are added to emulate the stiffening effect of ribs.  Introducing the change of coordinates $x'=x-L_f$, the panel with a rib at $x'=L_p/2$, is referred to as the \textit{rib-supported panel}, and the panel with ribs at $x'=L_p/4$ and $x'=3L_p/4$ is referred to as the \textit{double rib-supported panel}. The panel with the extra constraints are expected to excite a wider range of structural and thermal modes, which may pose as a challenge to the conventional aerothermal surrogates.

All the hypersonic ATE cases are computed using the HYPATE-X framework for up to 1 second, with a time step size of 1 ms, at the operating conditions: $M_\infty=7.523$, $P_\infty=3759.678$ Pa, $T_\infty=466.20$ K, which are the same as those used in the test dataset.  In total, six hypersonic ATE cases are considered to test the PIROMs, as listed in Table \ref{table: ate cases}, where N means that no constraints are applied.  For panels without ribs, the cases are labeled $XY$, where $X$ and $Y$ correspond to the boundary conditions at the leading and trailing edges, respectively. The panels with one and two rib supports are labeled as $XZY$ and $XMWY$, where $Z$, $M$, $W$ correspond to the boundary conditions at $x'=L_p/2$, $x'=L_p/4$ and $x'=3L_p/4$, respectively.  Note that the case $SS$ corresponds to the conventional simply-supported panel configuration.

\begin{table}
    \caption{Boundary conditions for the six ATE cases.}
    \label{table: ate cases}
    \begin{center}
        \begin{tabular}{c|ccccc}
            \hline
            Name & $x'=0$ & $x'=L_p/4$ & $x'=L_p/2$ & $x'=3L_p/4$ & $x'=L_p$ \\\hline
            $SS$   & S & N & N & N & S \\
            $CS$   & C & N & N & N & S \\
            $CC$   & C & N & N & N & C \\
            $CSC$  & C & N & S & N & C \\
            $CSSC$ & C & S & N & S & C \\\hline
            $CX$   & C & N & N & N & X \\
            \hline
        \end{tabular}
    \end{center}
\end{table}

\subsection{Baseline CFD-based Aerothermoelastic Results}\label{sec:effects of structural boundary conditions}

As a baseline study, the effects of structural boundary conditions on the transient ATE characteristics of the panel structure are explored using the CFD-based ATE solver using the first five cases in Table \ref{table: ate cases}.  The transient thermoelastic response of the $CS$ case is similar to the conventional $SS$ case, as shown in Figs. \ref{figs/sec5/cfd_thermoelastic_probed_without_ribs.png}-\ref{figs/sec5/cfd_aerothermal_without_ribs.png}.  In both cases, the panel deforms into flow with low-frequency structural oscillations induced by the slowly time-varying aerothermal loads, and the wall temperatures increase at similar rates.  However, the higher stiffness of the clamped BC in the $CS$ case suppresses the structural oscillation and reduces the maximum structural deformation.  In addition, the clamped leading edge results in a gradual change in the slope of deformation and significantly reduces the peaks in the aerothermal loads, when compared to the $SS$ case.

The $CC$ case is also shown in Figs. \ref{figs/sec5/cfd_thermoelastic_probed_without_ribs.png}-\ref{figs/sec5/cfd_aerothermal_without_ribs.png} and its ATE response is drastically different from the $SS$ and $CS$ cases.  The increased structural stiffness due to the clamped BC's results in two new characteristics in the thermoelastic response: 1) the panel thermally buckles at $\sim 0.35$ seconds, and 2) the panel deforms in the opposite direction of the $SS$ and $CS$ cases.  Moreover, the clamped BC's enforce zero slopes of deformation at the leading and trailing edges, and result in a more smooth distribution of aerothermal loads of lower amplitudes when compared to the previous two cases.

The ATE characteristics of the $CSC$ and $CSSC$ cases are similar to the $CC$ case, as shown in Figs. \ref{figs/sec5/cfd_thermoelastic_probing_M7523CCSSCC.png}-\ref{figs/sec5/cfd_aerothermal_with_ribs.png}, and the onset of thermal buckling is delayed further to $\sim 0.65$ s and $\sim 0.75$ s for the $CSC$ and $CSSC$ cases, respectively. In addition, due to the increased stiffness, the amplitudes of the distributions of deformation, temperature and aerothermal loads are all smaller than those in the cases without rib supports.  However, the rib supports induce more spatial variation in the structural deformation, leading to more nonlinear aerothermal load distributions and subsequently a highly non-uniform temperature distribution, which pose more challenges for the reduced-order aerothermal solver.

\begin{figure}
    \centering
    \insertfigs{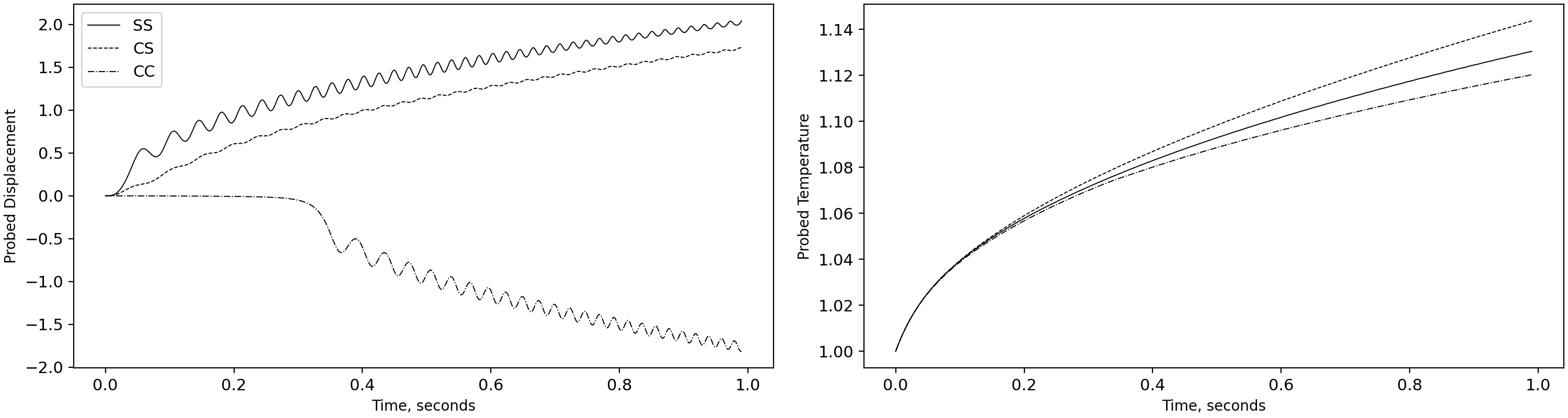}{0.92}{Structural and thermal responses at the center of the panel over time.}
    \insertfigs{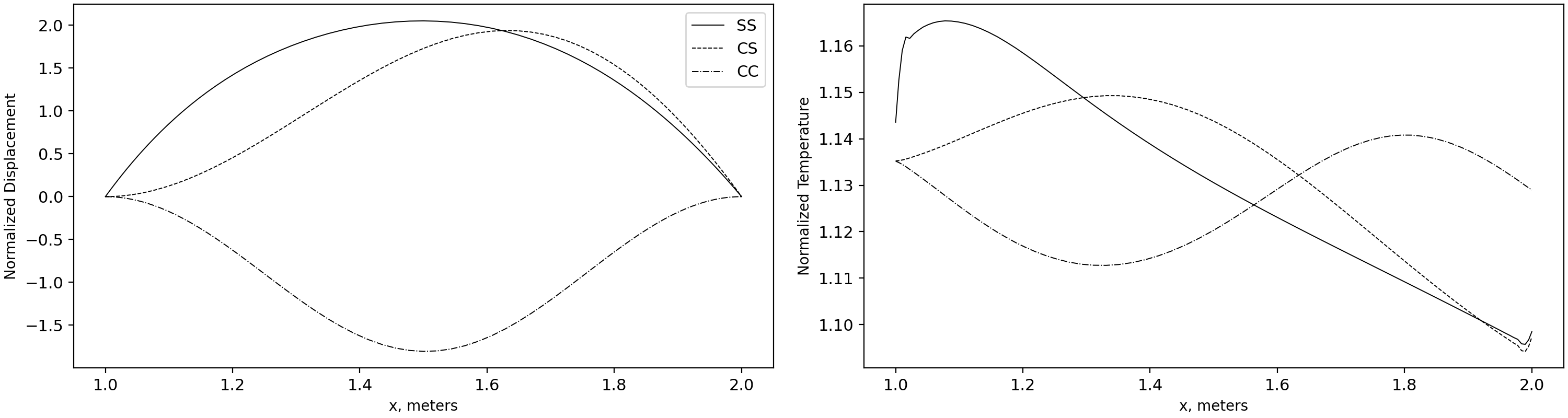}{0.92}{Structural deformation and temperature distribution at $t=1$s.}
    \insertfigs{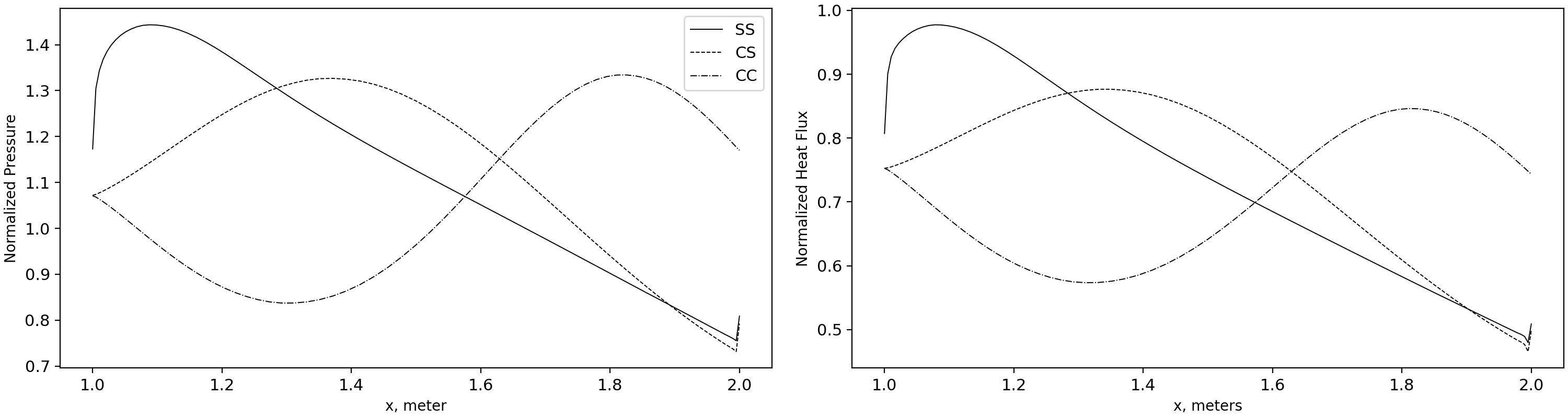}{0.92}{Distributions of pressure and heat flux at $t=1$s.}
    \caption{Thermoelastic and aerothermal responses for the $SS$, $CS$, and $CC$ cases.}
    \label{fig:cfd_without_ribs}
\end{figure}

\begin{figure}
    \centering
    \insertfigs{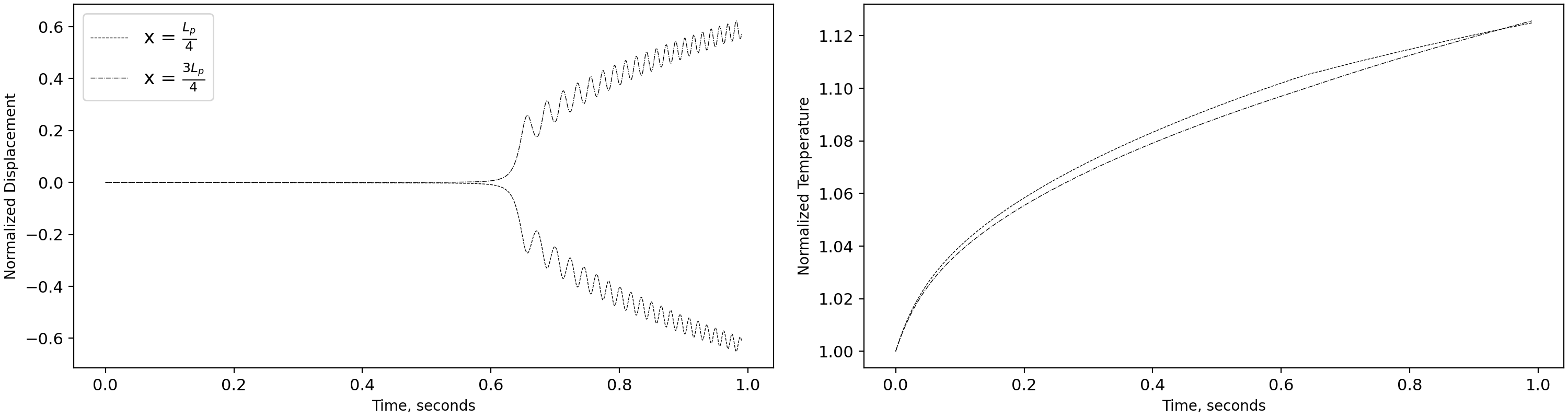}{0.92}{Structural and thermal responses for the $CSC$ case at $x'=L_p/4$ and $x'=3L_p/4$.}
    \insertfigs{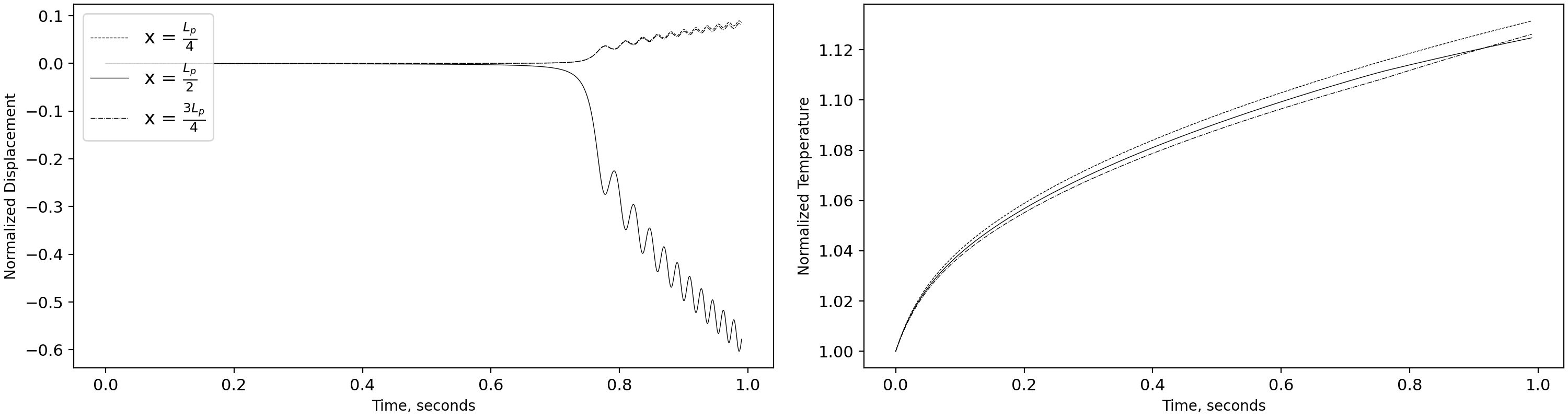}{0.92}{Structural and thermal responses for the $CSSC$ case at $x'=L_p/4$, $x'=L_p/2$ and $x'=3L_p/4$.}
    \insertfigs{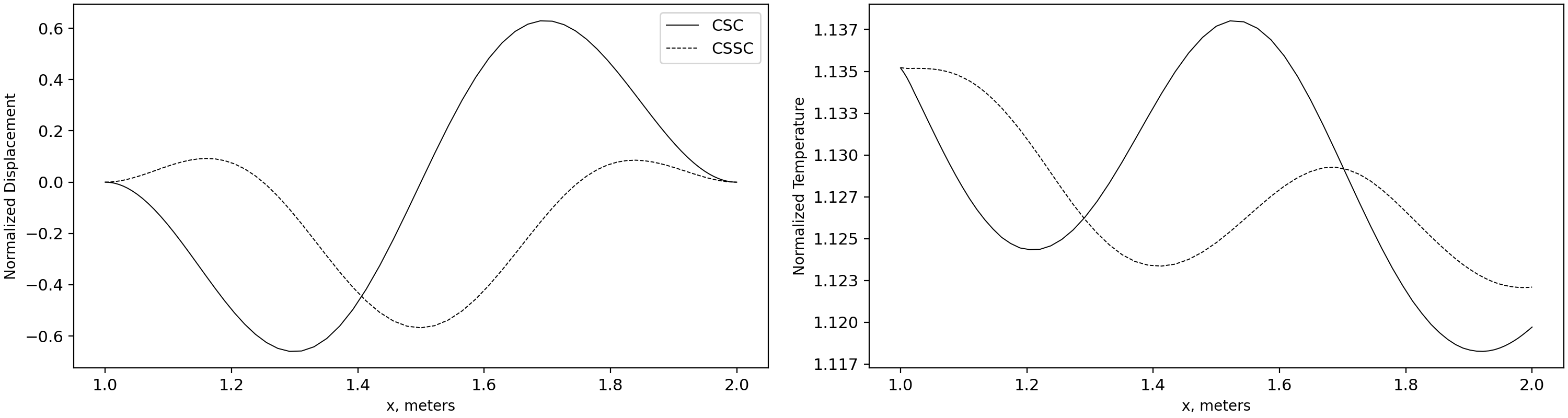}{0.92}{Structural deformation and temperature distribution at $t=1$s.}
    \insertfigs{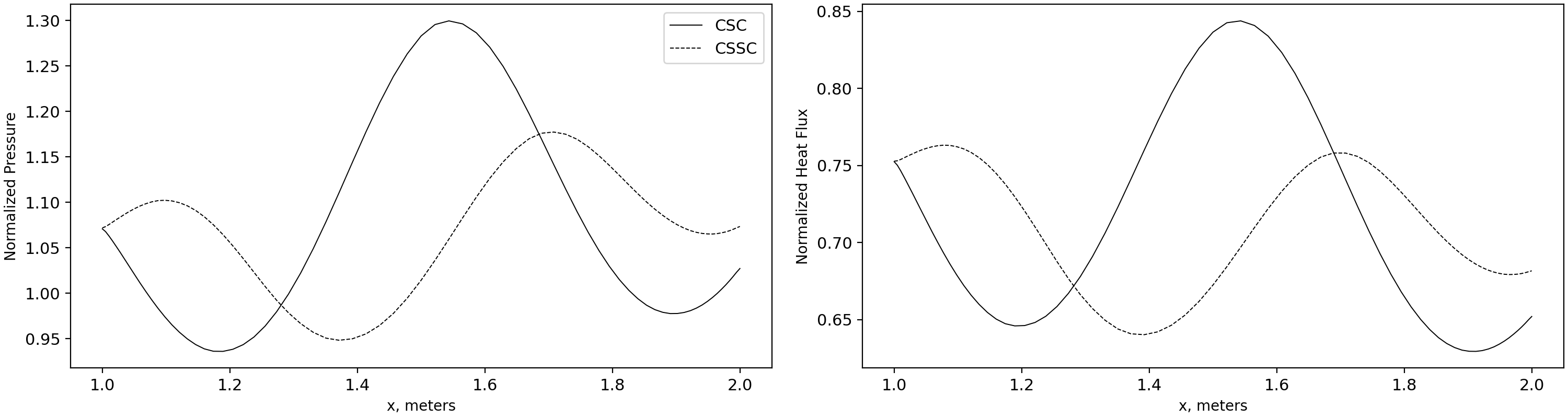}{0.92}{Distributions of pressure and heat flux at $t=1$s.}
    \caption{Thermoelastic and aerothermal responses for the $CSC$ and $CSSC$ cases.}
    \label{fig:cfd_with_ribs}
\end{figure}

\subsection{PIROM-based Aerothermoelastic Results}\label{sec: coupled results}

Next, the ATE simulation results based on the PIROM and POD-kriging models are compared against the CFD-based results for the different panel configurations.  The aerothermal ROM's only predict the quasi-steady component of the pressure caused by structural deformation, and the unsteady component, caused by the structural velocity, is accounted for using a piston-theory-based correction formulation \cite{Huang2019b}.

While the PIROM-d formulation shows some advantage in terms of accuracy over the PIROM-i formulation, PIROM-d experienced numerical instability and produced erroneous results when it is employed for the ATE simulation of the $SS$ case.  The reason is attributed to the abrupt change in the slope of deformation at the leading edge of the panel that caused excessive numerical stiffness in the solution of the ATVI equations.  Therefore, in the rest of this section, only the PIROM-i results are shown.

\begin{figure}
    \centering
    \insertfigs{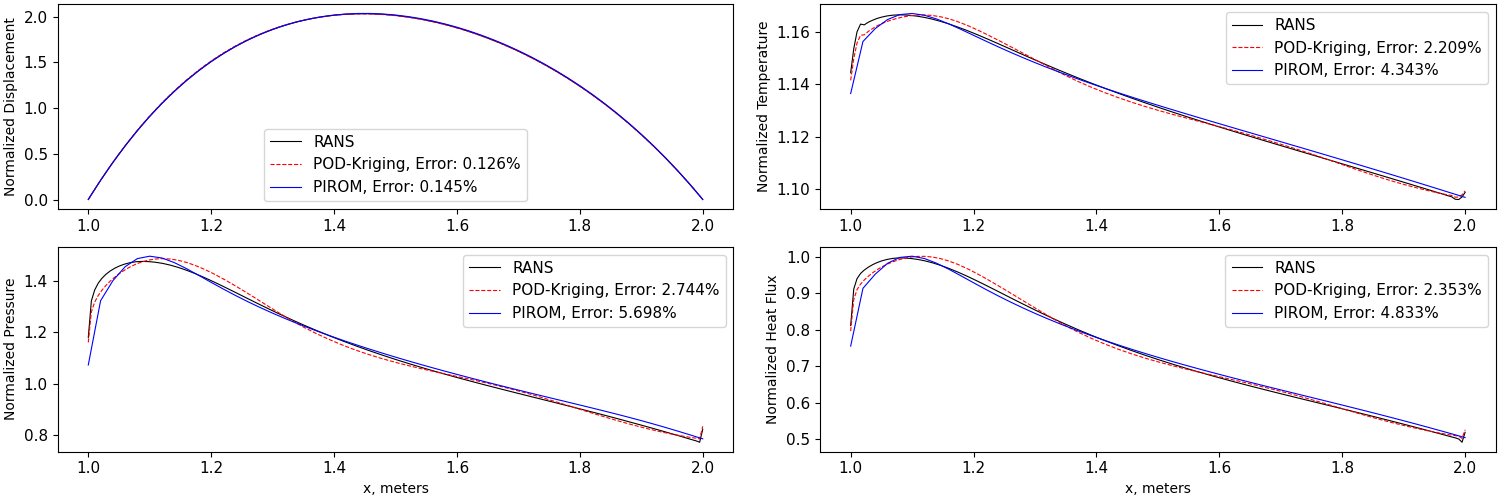}{0.92}{Case $SS$.}
    \insertfigs{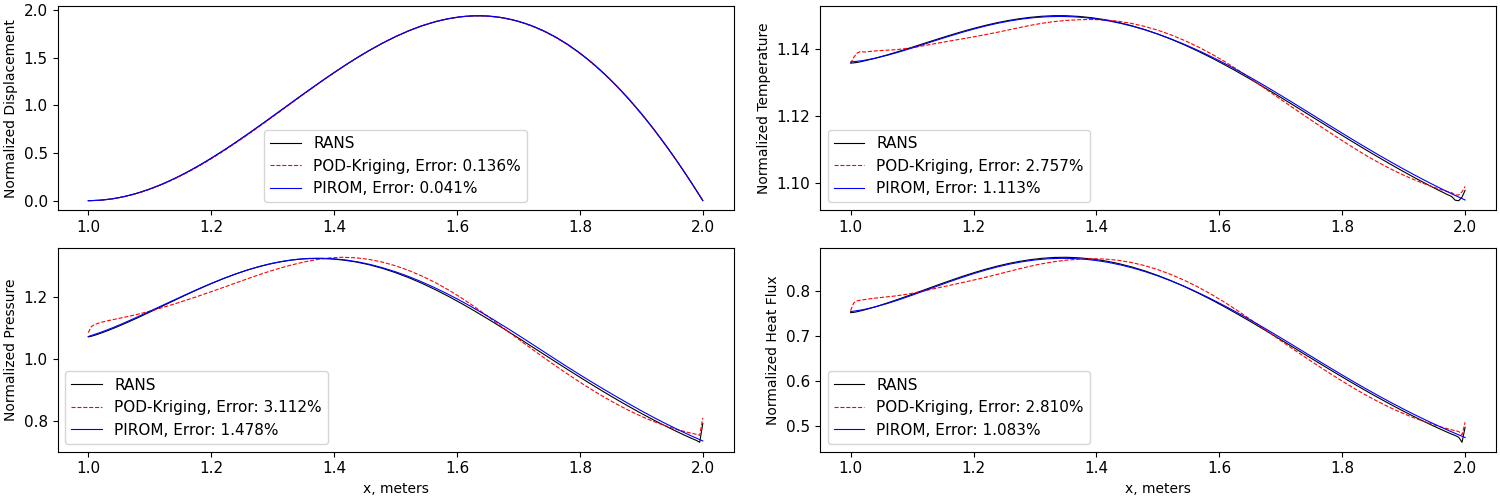}{0.92}{Case $CS$.}
    \insertfigs{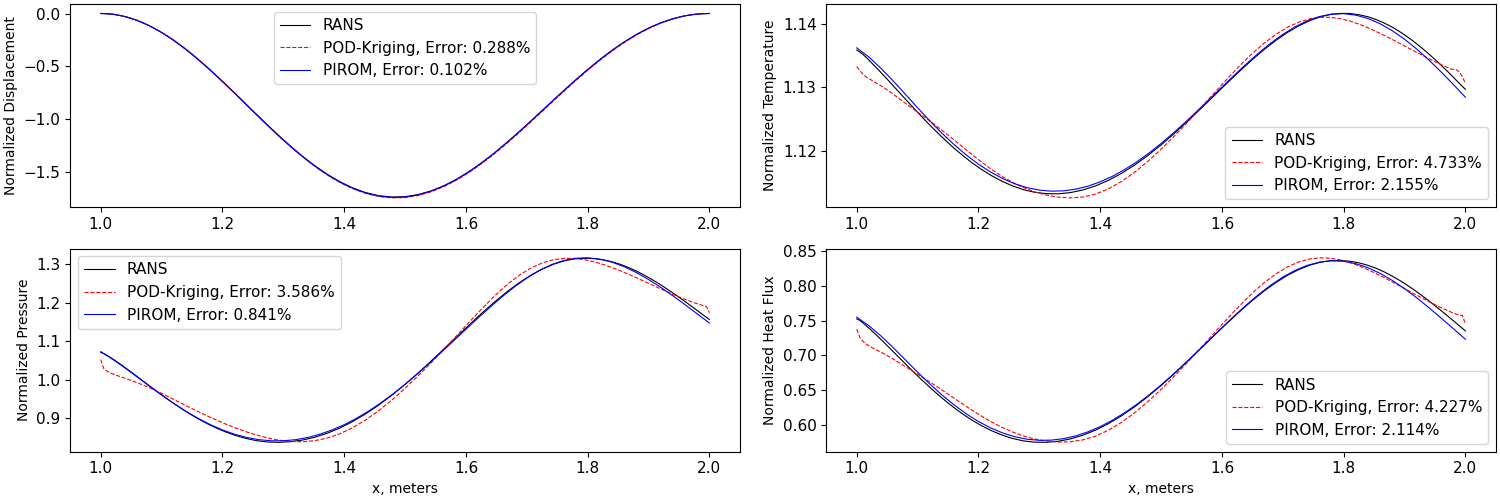}{0.92}{Case $CC$.}
    \caption{The ATE responses at $t=1.0$ s for the $SS$, $CS$ and $CC$ cases.}
    \label{fig_ate_rom_nr_rsp}
\end{figure}

The ROM-based ATE responses for the $SS$, $CS$ and $CC$ cases are compared against the CFD-based results in Fig. \ref{fig_ate_rom_nr_rsp}.  Both models perform reasonably well with errors less than 5\% for almost all quantities of interest and POD-kriging outperforms PIROM only in the $SS$ case.  The high accuracy of POD-kriging is expected for the $SS$ case since the structure is subjected to the same operating conditions and BC's used in the training samples for this model.

\begin{figure}
    \centering
    \insertfigs{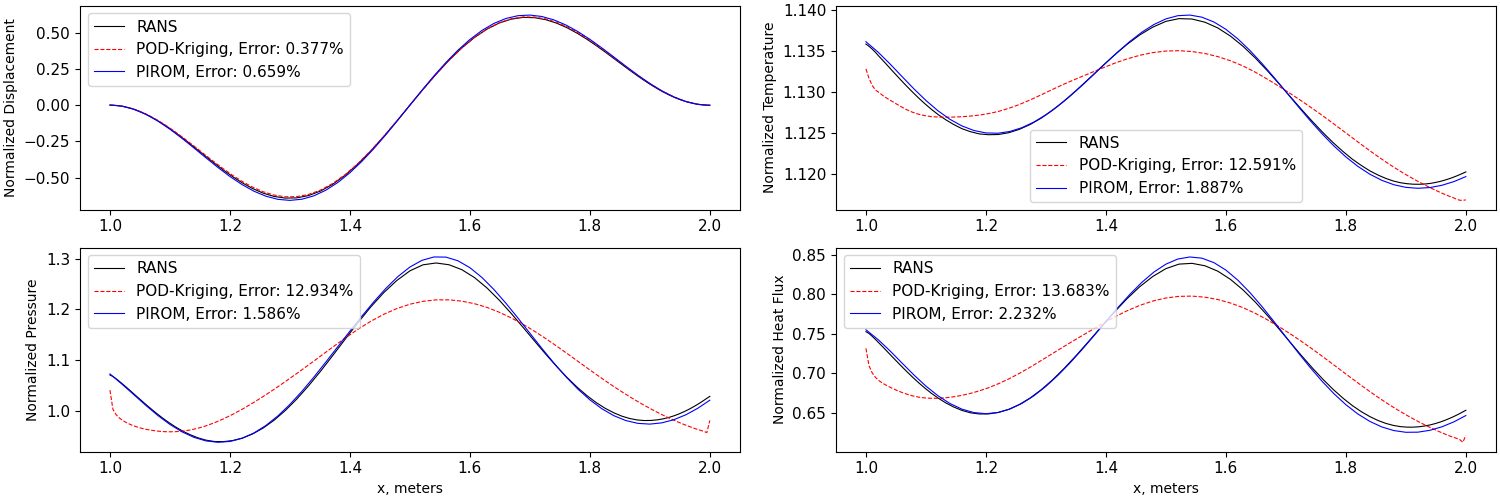}{0.92}{Case $CSC$.}
    \insertfigs{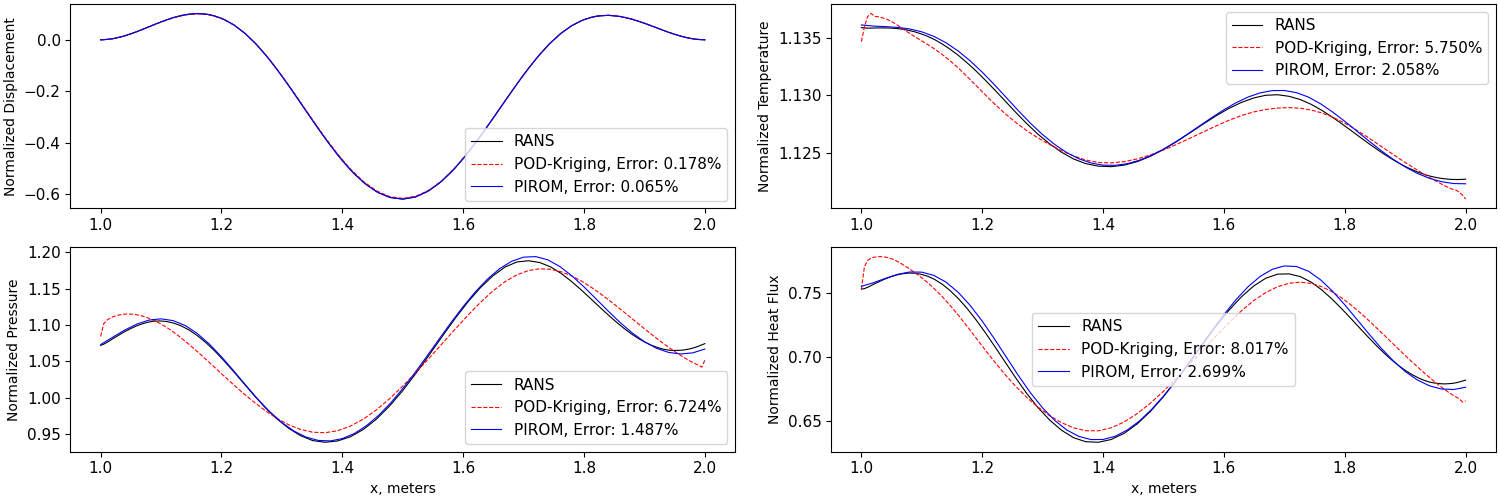}{0.92}{Case $CSSC$.}
    \caption{The ATE responses at $t=1.0$ s for the $CSC$ and $CSSC$ cases.}
    \label{fig_ate_rom_rb_rsp}
\end{figure}

\begin{figure}
    \centering
    \insertfigs{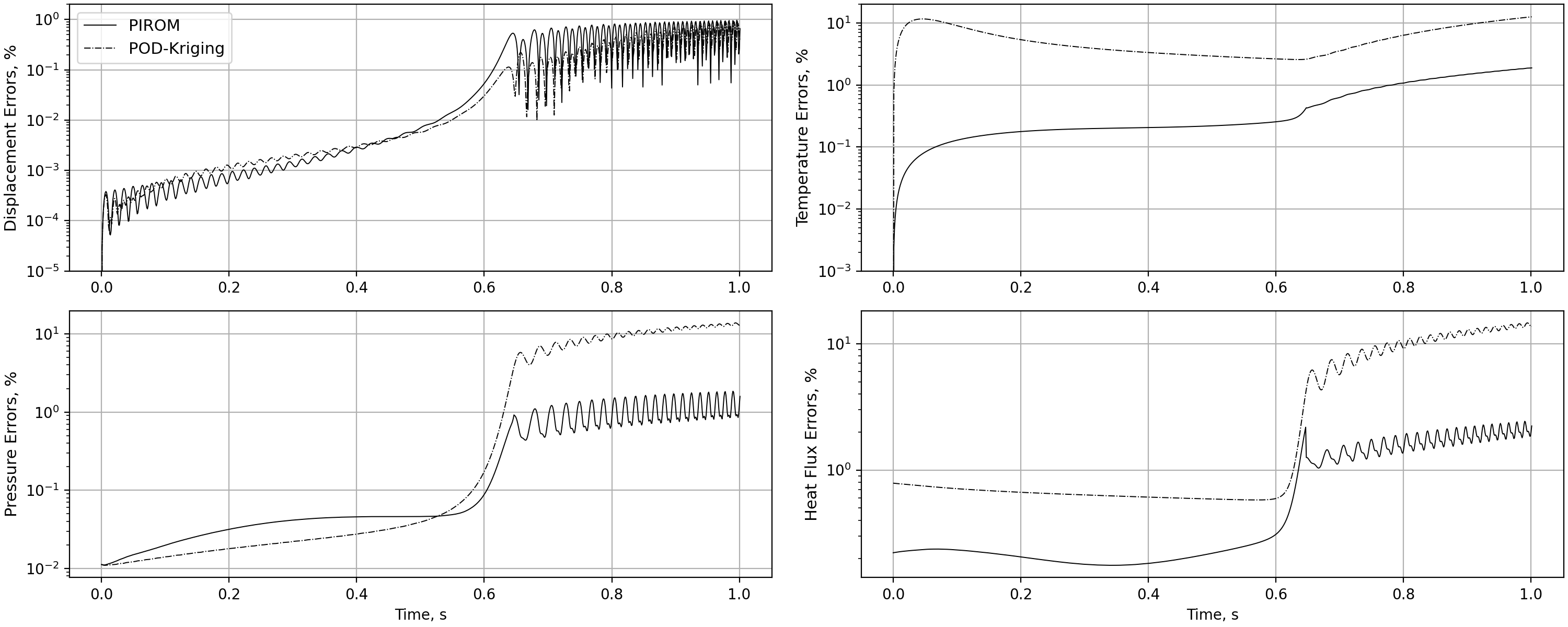}{0.92}{Case $CSC$.}
    \insertfigs{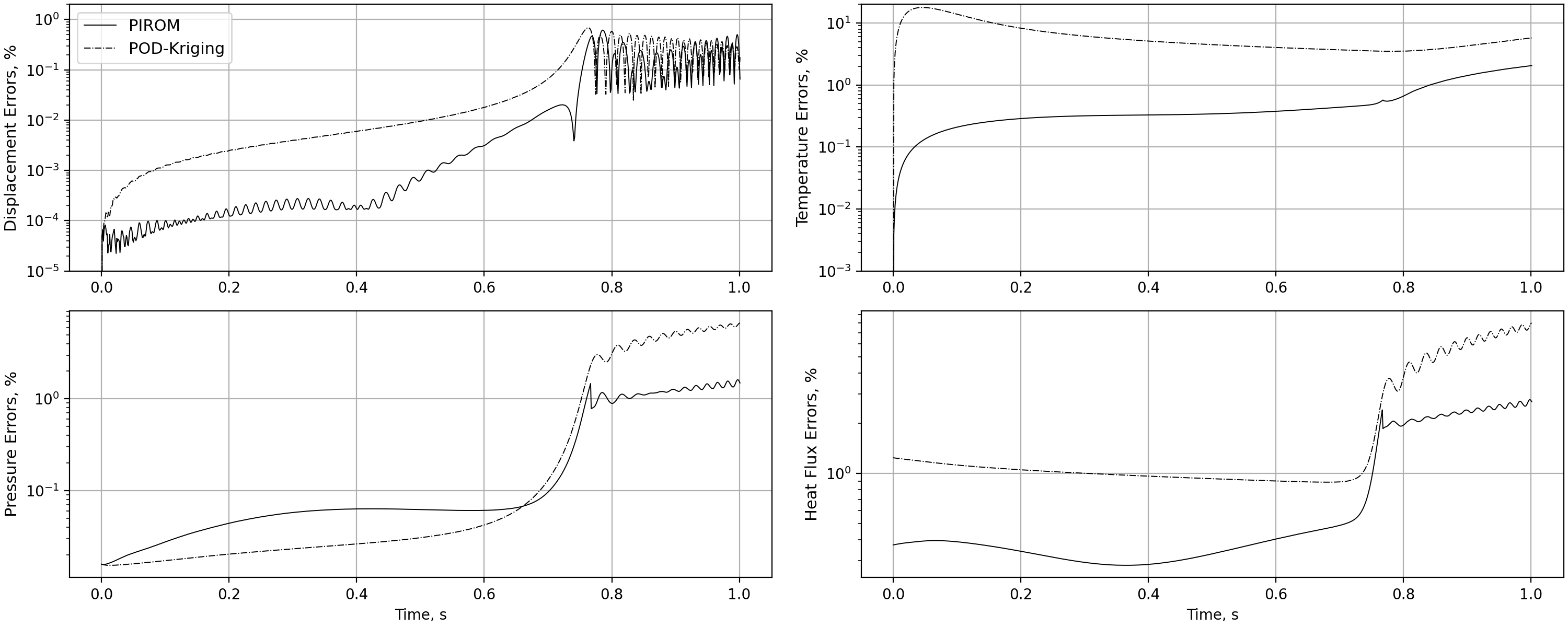}{0.92}{Case $CSSC$.}
    \caption{The NRMSE error over time for the ROM-based ATE responses for the $CSC$ and $CSSC$ cases.}
    \label{fig_ate_rom_rb_err}
\end{figure}

Next, the results for the $CSC$ and $CSSC$ cases are shown in Fig. \ref{fig_ate_rom_rb_rsp}, where the superiority of PIROM over the POD-kriging surrogate becomes more apparent.  The aerothermal load predictions from PIROM show a significant improvement relative to the POD-kriging predictions, in terms of the magnitudes and locations of the maxima and minima.  The improvement is particularly evident for the $CSC$ case\CORRECTa{, since the structural response in this case deviates more from the sinusoidal mode shapes used in the POD-kriging model}.  In addition, the errors over time in the ROM-based ATE responses for the $CSC$ and $CSSC$ cases are shown in Fig. \ref{fig_ate_rom_rb_err}.  For the POD-kriging-based results, the errors quickly accumulate after the panel buckles and exceed 10\% at the end of simulation in some cases.  On the contrary, the PIROM produces a relatively more accurate aerothermal prediction so that the errors in the PIROM-based ATE response are maintained to be only $\sim$1\%.

A curious observation in Figs. \ref{fig_ate_rom_rb_rsp} and \ref{fig_ate_rom_rb_err}, however, is that the errors in the structural response are extremely low and less than 0.3\% for both the PIROM and POD-kriging results, even if the error in POD-kriging prediction is up to over 13\%.  The explanation is that, in the current setup, the structural deformation is mainly driven by the average thermal stress and is relatively insensitive to the temperature \textit{distribution}.  Furthermore, as its magnitude increases, the deformation becomes even less sensitive to the average thermal stress due to the increased geometrical stiffness effect.  Nevertheless, the accurate aerothermal prediction is still important.  The errors in the predicted heat flux distribution are clearly correlated to the errors in the thermal responses.  For hypersonic structures operating under a high-temperature environment, the accurate prediction of the wall temperature distribution, including the magnitudes and locations of its peaks, is critical for the characterization of the service life of the structural material.

\subsection{Effects of Spring-Constrained Boundary Conditions}\label{sec: parametric study}

The simply-supported and clamped BC's considered so far are both structural idealizations and unlikely to occur in real structures.  The BC's at the leading and trailing edges are better characterized by a spring-constrained BC, i.e., a simply-supported BC with a torsional spring constraint.  It is well known that the spring constant $k$ significantly alters the structural modal properties, as illustrated in Fig. \ref{figs/sec5/ss2cc_frequencies.png}, where the modal frequencies are normalized by those of the $SS$ case.  Between the blue and green curves, the $SS$ case transitions to the $CS$ case with increasing $k$ at the leading edge, while between the green and red curves the $CS$ case transitions to the $CC$ case with increasing $k$ at the trailing edge.

From the ATE responses in the $CS$ and $CC$ cases presented in the previous sections, it is clear that the choice of BC's has a dramatic effect on the transient thermoelastic characteristics of the panel.  Switching from a simply-supported BC to a clamped one causes the panel to deform out of the flow instead of into the flow.  In view of the $CS$ and $CC$ results, as the spring constant increases, a drastic transition in the structural response of the panel is expected.

The effects of the spring constant are explored via a parametric study consisting of 8 PIROM-based ATE simulations for the $CX$ configuration.  Figure \ref{figs/sec5/disp_vs_freq.png} shows the center-panel displacements averaged over the last 50 time steps versus the relative frequency.  Also, as a reference, the simulations based on CFD, POD-kriging, and TVI models are also included.  Despite the sensitivity of the ATE response to the spring constant, the PIROM prediction almost \textit{exactly} reproduces the CFD-based results, with a relative frequency of $2.255$ at the transition, which correspond to to a spring constant of $\sim8500$ N/m. \CORRECTa{The TVI and POD-kriging models predict lower transition spring constants with the values of $\sim6500$ N/m and $\sim7750$ N/m, which translate to errors relative to the CFD-based predictions of $23.529\%$ and $8.823\%$, respectively.}


The errors in ATE responses over time from the PIROM, POD-kriging, and TVI results are compared in Fig. \ref{figs/sec5/spring_mean_errors.png}, where the prediction errors in time are averaged among the 8 simulations.  Likely due to the complexities in the boundary conditions, both the pure physics-based model, TVI, and the pure data-driven model, POD-kriging, produce errors on the order of 10\%.  However, the PIROM results consistently maintained errors on the order of 1\%, which again highlights its accuracy and robustness in coupled ATE analysis under complex operating and boundary conditions.

\insertfig{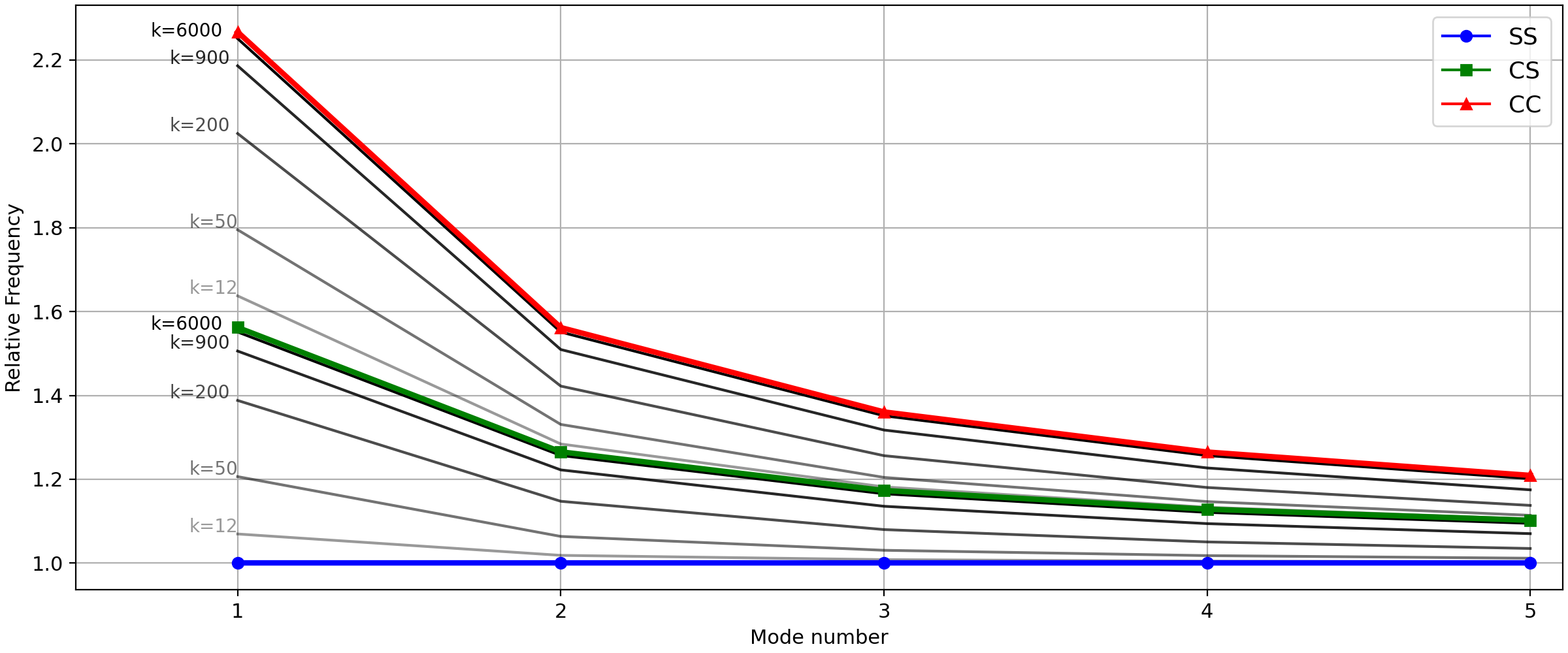}{0.80}{Variation of structural modal frequencies with increasing spring constant from $SS$ to $CS$ and $CC$ configurations.}

\begin{figure}
    \centering
    \insertfigs{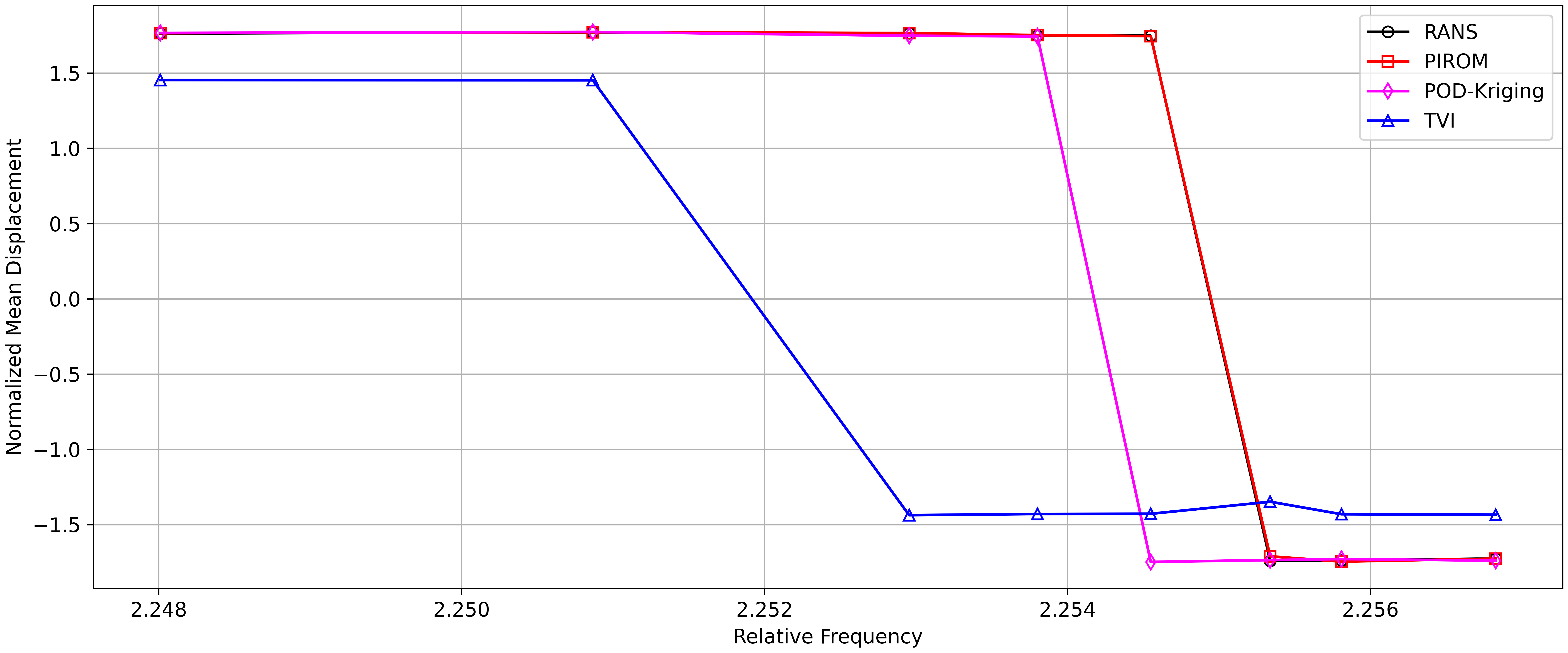}{0.80}{The center-panel displacement versus normalized modal frequency for the CX case.}
    \insertfigs{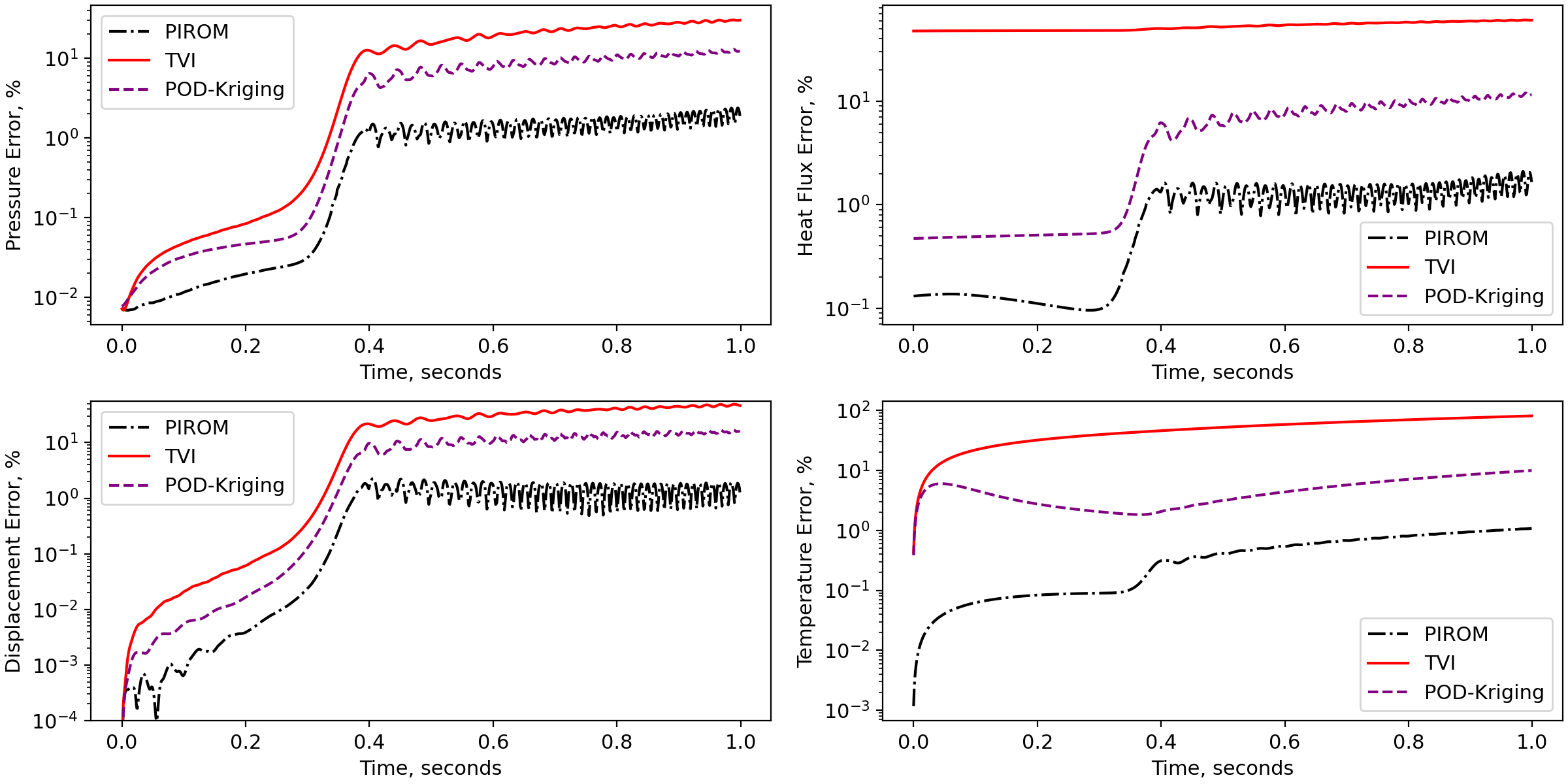}{0.80}{Comparison of mean errors over time in the TVI, PIROM, and POD-kriging results for the CX case.}
    \caption{Parametric study on the spring constant for the $CX$ case.}
    \label{fig: frequency analysis}
\end{figure}

\subsection{Computational Cost and Comparison of PIROM Models}

Lastly, the computation costs for generating and applying the aerothermal models are examined.  Obtaining an aerothermal load prediction from the PIROM-i, PIROM-d or POD-kriging model requires the following stages: 1) sample generation and simulation, 2) optimization and training, and 3) evaluation.  Stages 1 and 2 correspond to the one-time offline costs while Stage 3 corresponds to the online cost and is the most relevant for computational efficiency of coupled ATE simulation.  The detailed comparison of the three stages are shown in Table \ref{table:time}.  All computations are done on a workstation with Intel\textcopyright Xeon Silver 4214 processors.  The CFD simulations are performed in parallel with 10 cores while all other computations are performed in a serial manner.

Stage 1 involves sampling the parameter space and generating the high-fidelity CFD-based solutions.  Each sample requires executing a steady RANS simulation to obtain the high-fidelity aerothermal load solution, which requires approximately $160$ seconds of computing time. The PIROMs introduce a significant computational advantage in the process of generating the high-fidelity data, since 500 simulations are required for POD-kriging, whereas the PIROMs require 150 simulations.
The analysis of stage 2 is more involved.  Training the POD-kriging surrogate simply requires applying the POD and GP to the 500 training samples, which translates to a training time of approximately 0.5 hours for the pressure and heat flux data together.  The training of PIROM-i requires first 150 optimizations to obtain the optimal augmentation values, which results in approximately 45 hours of computing time, and then the fitting of the augmentation values using the GP models.  Due to the large amount data, the GP fitting consumes 6 hours in total.  The PIROM-d circumvents the need to generate the training data, however, the depth, stiffness and nonlinearity of the NODEs requires 16 hours of training time.
Finally, Stage 3 is concerned with the computing time required by the aerodynamic solvers to provide an aerothermal load solution to the structural and thermal solvers in HYPATE-X to advance the simulation in time.  All the ROM's are 10-100 times faster than the CFD solver, even though run in a serial setting.  The costs for PIROM-i and POD-kriging are on the same order of magnitude, whereas the PIROM-d is one order of magnitude slower due to the numerical stiffness in NODE.

Comparing the computational costs among the aerothermal ROM's, the accuracy, robustness, and generalization capabilities of PIROM comes at the expense increased offline computing time to create the aerothermal model. \CORRECTa{Among the PIROMs, the PIROM-d reduces the computational cost by removing the optimization phase in PIROM-i.  In addition, PIROM-d scales better than PIROM-i for large training datasets, as the latter requires a non-parametric model such as GP as the data-driven component.  The computational cost for training GP models scales cubically with number of training samples, and thus PIROM-i may face a bottleneck in computation when more training samples are considered.  To sum, there is a trade-off between the numerical robustness and computational cost between PIROM-i and PIROM-d, and the readers are recommended to select the appropriate formulations depending on their applications.} 

\CORRECTa{The comparison in this section also indicates that more} algorithmic development to enhance the training efficiency \CORRECTa{of PIROM} is required as future work.  In addition, the numerical stability of PIROM-d also needs to be enhanced in order to be robustly incorporated into the ATE simulation.

\begin{table}
    \caption{Comparison of computational costs for the aerothermal models.}
    \label{table:time}
    \begin{center}
    \resizebox{\textwidth}{!}{
        \begin{tabular}{*{5}{c}}
            \hline
            Model       & Sample Generation (hours) & Optimization (hours) & Training (hours) & Prediction (seconds) \\\hline
            RANS        & -                         & -                    & -                & 44.6 \\
            POD-kriging & 22.2                    & -                    & 0.5              & 0.137 \\
            PIROM-i     & 6.67                     & 45                   & 6                & 0.485 \\
            PIROM-d     & 6.67                     & -                    & 16               & 4.86 \\
            \hline
        \end{tabular}}
    \end{center}
\end{table}

\section{Conclusions}\label{sec:conclusion}

In this study, the physics-infused reduced-order modeling (PIROM) methodology is presented and applied to aerothermal load modeling for hypersonic aerothermoelastic analysis. \CORRECTa{As a general reduced-order modeling} methodology, PIROM explicitly couples a \CORRECTa{first-order} physics-based model with a data-driven component through either an algebraic or differential augmentation formulation. 

The \CORRECTa{PIROM-based aerothermal model} overcomes the characterization issue and significantly relieves the generalization issue in the conventional aerothermal surrogates, represented by the POD-kriging method.  The physics-based component of PIROM allows for arbitrary operating conditions and continuous thermoelastic inputs, and thus no longer requires a parametrization of the inputs (i.e., characterization).  The physics-based component also provides the prior knowledge on the trend of the model output over the entire input space of interest, and thus facilitates the generalizability of the model.  Furthermore, the data-driven augmentation component enhances the predictive accuracy of the aerothermal model by correcting the approximations and semi-empirical expressions in the physics-based component.  The augmentation formulation allows for a clear physical interpretation of the data-driven components, which is missing in many purely data-driven models.  Finally, training algorithms are developed to learn the PIROM from data, including the indirect approach based on the FIML method that are suitable for algebraic augmentation, and the direct approach that are extended to train PIROM with differential augmentation.


The specific results and findings are listed as follows,
\begin{compactenum}
    \item A systematic benchmark of the PIROM-based aerothermal model is performed against high-fidelity CFD solutions and a conventional POD-kriging-based model generated for a simply-supported panel, via (1) test datasets of synthesized thermoelastic inputs and (2) fully-coupled transient ATE analysis.
    \begin{compactenum}
        \item In all cases PIROM outperforms the POD-kriging method in terms of accuracy, except in the simply-supported case where the two models show comparable accuracy.  The comparison verifies the accuracy and generalizability of the PIROM in the aerothermal applications.
        \item The PIROM-based aerothermal solver performs robustly and accurately in prolonged hypersonic ATE simulations for structures with complex boundary conditions and constraint, and the errors in ATE responses are maintained on the order of $\sim$1\%.  In contrast, the errors in POD-kriging-based results are as high as 6-13\%.
        \item The PIROM has a high sampling efficiency that exploits the spatial resolution of the high-fidelity sample solution, so that only $10$-$10^2$ samples are sufficient for model training, in contrast to $10^2$-$10^3$ samples for conventional interpolation-based models.
        \item The online prediction costs for PIROM-i and POD-kriging are on the same order of magnitude, whereas the PIROM-d is slower due to the numerical stiffness issue.  However, the enhanced capability of PIROM does come at the price of increased computational cost in the offline training, when compared to the POD-kriging.
    \end{compactenum}
    \item The effects of structural boundary conditions and constraints on the ATE responses are examined, including simply-support, clamped, and spring-constraint for the leading and trailing edges, as well as the rib-supports along the panel that are emulated by simply-supported constraints.
    \begin{compactenum}
        \item The clamped and rib-supported configurations delay the onset of thermal buckling relative to the simply-supported configurations, and result in distributions of deformation and temperature that have more spatial variation but smaller amplitudes.
        \item In all ATE cases, the errors in aerothermal prediction propagates to the thermal response but do not have a strong effect on the structural response.
        \item When a relatively realistic spring-constrained BC at the leading and trailing edges is considered, the ATE response is sensitive to the stiffness characterization of the BC's.  Increasing the spring constant of the trailing edge BC, the panel deformation may drastically transition from the into-flow direction to out-of-flow direction, and hence change the distributions of aerothermal loads and the temperature distributions.
    \end{compactenum}
\end{compactenum}

In sum, the above results demonstrates that the PIROM methodology can be employed to generate an accurate, efficient and robust aerothermal model for coupled ATE analysis of complex structural configurations, without needing to parametrize the configuration geometry.  This unique feature makes the novel PIROM methodology a promising tool to facilitate rapid yet accurate aerothermoelastic design and optimization of hypersonic structures in a dynamical setting.

\CORRECTa{Furthermore, the initial success establishes PIROM as a new reduced-order modeling methodology, and builds up the confidence in the extension of PIROM-based aerothermal models to more complex engineering problems such as aerothermoelasticity of scramjets and inlets involving shock wave-boundary layer interactions, which features localized aerothermal heating and possibly small-scale surface bumps.  Currently such problems are tackled using methods similar to POD or POD-kriging \cite{Xiaoxuan2019,Cottier2019,Xiaoxuan2020}.  It is possible to apply PIROM to extend the classical triple deck theory \cite{Smith1982,Ryzhov2012} to produce a new aerothermal model with improved generalizability.}

Nevertheless, some challenges still remain to be tackled towards an improved and more numerically robust PIROM methodology.  The most significant limitation in PIROM originates from the numerical stiffness in PIROM-d, and proper numerical conditioning of PIROM-d shall be investigated in the future.

\appendix
\section{Definition of Thermoelastic Modes for Training}\label{appendix:coefficients}
The structural and thermal modes used in training of the PIROMs are defined as $11^{th}$ order polynomials. Polynomials are chosen for easier computation of derivatives of the structural and temperature distributions, which are required in the PIROM evaluation. The polynomial coefficients for the structural modes are given in Table \ref{table: structure coefficients} , and for the thermal modes in Table \ref{table: thermal coefficients}

\begin{table}[H]
    \caption{Structural mode polynomial coefficients.}
    \label{table: structure coefficients}
    \begin{center}
    \resizebox{0.62\columnwidth}{!}{
        \begin{tabular}{|c|c|c|c|c|}
            \hline
            Order  & D1 & D2 & D3 & D4 \\\hline
            $x^{11}$  & $-3.939\times10^{0}$ & $1.660\times10^{-2}$ & $-8.354\times10^{-3}$ & $-4.789\times10^{0}$ \\
            $x^{10}$ & $-3.080\times10^{0}$ & $-2.446\times10^{-4}$ & $-1.693\times10^{-1}$ & $-1.171\times10^{0}$ \\
            $x^9$  & $8.289\times10^{-1}$ & $-1.024\times10^{-1}$ & $-7.626\times10^{-2}$ & $8.190\times10^{0}$ \\
            $x^8$  & $2.550\times10^{0}$ & $2.352\times10^{-3}$ & $3.873\times10^{-1}$ & $1.294\times10^{0}$ \\
            $x^7$  & $1.526\times10^{0}$ & $3.125\times10^{-1}$ & $1.625\times10^{-1}$ & $-7.118\times10^{0}$ \\
            $x^6$  & $2.205\times10^{-2}$ & $-1.337\times10^{-1}$ & $-4.866\times10^{-1}$ & $-6.171\times10^{-1}$ \\
            $x^5$  & $-2.535\times10^{-1}$ & $-3.800\times10^{-1}$ & $-1.282\times10^{-1}$ & $3.526\times10^{0}$ \\
            $x^4$  & $-1.914\times10^{-1}$ & $4.063\times10^{-2}$ & $3.371\times10^{-1}$ & $1.386\times10^{-1}$ \\
            $x^3$  & $-1.343\times10^{-1}$ & $7.700\times10^{-2}$ & $3.854\times10^{-2}$ & $-8.469\times10^{-1}$ \\
            $x^2$  & $-7.538\times10^{-3}$ & $-4.940\times10^{-2}$ & $-9.822\times10^{-2}$ & $-1.175\times10^{-2}$ \\
            $x^1$  & $2.620\times10^{-2}$ & $1.147\times10^{-9}$ & $-3.858\times10^{-3}$ & $7.522\times10^{-2}$ \\
            $x^0$  & $6.550\times10^{-3}$ & $1.001\times10^{-2}$ & $9.739\times10^{-3}$ & $-8.910\times10^{-7}$ \\
            \hline
        \end{tabular}}
    \end{center}
\end{table}

\begin{table}[H]
    \caption{Thermal mode polynomial coefficients.}
    \label{table: thermal coefficients}
    \begin{center}
    \resizebox{0.78\columnwidth}{!}{
        \begin{tabular}{|c|c|c|c|c|c|}
            \hline
            Order & T1 & T2 & T3 & T4 & T5 \\\hline
            $x^{11}$& $-1.068\times10^{-7}$ & $1.522\times10^{-7}$ & $1.879\times10^{3}$ & $1.933\times10^{2}$ & $6.381\times10^{4}$ \\
            $x^{10}$& $-6.091\times10^{-8}$ & $-5.399\times10^{-8}$ & $-2.806\times10^{4}$ & $2.869\times10^{1}$ & $-3.427\times10^{3}$ \\
            $x^9$& $1.036\times10^{4}$ & $-3.659\times10^{2}$ & $-1.403\times10^{2}$ & $-1.224\times10^{2}$ & $-3.238\times10^{4}$ \\
            $x^8$& $2.922\times10^{3}$ & $3.255\times10^{2}$ & $6.668\times10^{3}$ & $-1.377\times10^{1}$ & $5.461\times10^{3}$ \\
            $x^7$& $-5.076\times10^{3}$ & $3.767\times10^{2}$ & $4.244\times10^{3}$ & $2.707\times10^{1}$ & $4.319\times10^{3}$ \\
            $x^6$& $-1.335\times10^{3}$ & $2.260\times10^{2}$ & $4.314\times10^{2}$ & $2.582\times10^{0}$ & $-3.265\times10^{3}$ \\
            $x^5$& $7.063\times10^{2}$ & $3.852\times10^{1}$ & $-3.782\times10^{2}$ & $-2.679\times10^{0}$ & $3.682\times10^{2}$ \\
            $x^4$& $2.913\times10^{2}$ & $-9.581\times10^{1}$ & $-2.657\times10^{2}$ & $1.158\times10^{-1}$ & $9.768\times10^{2}$ \\
            $x^3$& $1.384\times10^{2}$ & $-4.106\times10^{1}$ & $-1.587\times10^{2}$ & $-3.794\times10^{-1}$ & $-1.589\times10^{2}$ \\
            $x^2$& $-7.949\times10^{1}$ & $-1.426\times10^{1}$ & $5.971\times10^{1}$ & $7.279\times10^{-1}$ & $-1.277\times10^{2}$ \\
            $x^1$& $-2.364\times10^{1}$ & $-4.308\times10^{0}$ & $3.149\times10^{1}$ & $-2.033\times10^{0}$ & $8.280\times10^{0}$ \\
            $x^0$ & $5.87\times10^{1}$ & $5.12\times10^{1}$ & $4.70\times10^{1}$ & $5.20\times10^{1}$ & $5.25\times10^{1}$ \\
            \hline
        \end{tabular}}
    \end{center}
\end{table}
\clearpage

\bibliography{references}

\end{document}